\documentclass[preprint2]{aastex6}

\usepackage{amsmath}

\newcommand{\HII}{H$\,${\sc ii}}

\newcommand{\kms}{km\,s$^{-1}$}

\def\ts   {\thinspace}
\def\kms  {\ifmmode{{\rm \ts km\ts s}^{-1}}\else{\ts km\ts s$^{-1}$}\fi}
\def\mo   {\ifmmode{{\rm M}_{\odot}}\else{M$_{\odot}$}\fi}

\def\aco {\ifmmode{^{12}{\rm CO}(J=1\to0)}\else{$^{12}{\rm
CO}(J=1\to0)$}\fi}
\def\bco {\ifmmode{^{12}{\rm CO}(J=2\to1)}\else{$^{12}{\rm
CO}(J=2\to1)$}\fi}
\def\m  {\ifmmode{\mu {\rm m}}\else{$\mu$m}\fi}
\def\cco {\ifmmode{^{13}{\rm CO}(J=1\to0)}\else{$^{13}{\rm
CO}(J=1\to0)$}\fi}
\def\dco {\ifmmode{^{13}{\rm CO}(J=2\to1)}\else{$^{13}{\rm
CO}(J=2\to1)$}\fi}
\def\eco {\ifmmode{{\rm C}^{18}{\rm O}(J=1\to0)}\else{{\rm C}$^{18}{\rm
O}(J=1\to0)$}\fi}
\def\hi  {\ifmmode{{\rm H}{\rm \small I}}\else{H\ts {\scriptsize I}}\fi}
\def\hii  {\ifmmode{{\rm H}{\rm \small II}}\else{H\ts {\scriptsize II}}\fi}
\def\Ha  {\ifmmode{{\rm H}{\alpha}}\else{H\ts {$\alpha$}}\fi}
\def\Hb  {\ifmmode{{\rm H}{\alpha}}\else{H\ts {$\beta$}}\fi}

\def\nh  {\ifmmode{N(\hi)}\else{$N$(\hi)}\fi}
\def\hun  {\ifmmode{I_{100}}\else{$I_{100}$}\fi}
\def\sex  {\ifmmode{I_{60}}\else{$I_{60}$}\fi}
\def\hh   {\ifmmode{{\rm H}_2}\else{H$_2$}\fi}
\def\nhh   {\ifmmode{N({\rm H}_2)}\else{$N$(H$_2$)}\fi}
\def\zwco  {\ifmmode{^{12}{\rm CO}}\else{$^{12}{\rm CO}$}\fi}
\def\nzwco  {\ifmmode{N(^{12}{\rm CO})}\else{$N(^{12}{\rm CO})$}\fi}
\def\wzwco  {\ifmmode{W(^{12}{\rm CO})}\else{$W(^{12}{\rm CO})$}\fi}
\def\drco  {\ifmmode{^{13}{\rm CO}}\else{$^{13}{\rm CO}$}\fi}
\def\ndrco  {\ifmmode{N(^{13}{\rm CO})}\else{$N(^{13}{\rm CO})$}\fi}
\def\wdrco  {\ifmmode{W(^{13}{\rm CO})}\else{$W(^{13}{\rm CO})$}\fi}
\def\tex  {\ifmmode{T_{ex}({\rm CO})}\else{$T_{ex}({\rm CO})$}\fi}
\def\xco   {\ifmmode{X_{\rm CO}}\else{$X_{\rm CO}$}\fi}
\def\msol   {\ifmmode{{\rm M}_{\odot}}\else{M$_{\odot}$}\fi}
\def\amm    {NH$_{3}$}
\def\mh     {H$_{2}$}

\def\methanol {CH$_3$OH}
\def\water {H$_2$O} 
\def\sfr {M$_{\odot}$\,yr$^{-1}$}
\def\Lsun{L$_{\odot}$}

\begin{document}

\title{Survey of Water and Ammonia in Nearby galaxies (SWAN): Resolved Ammonia Thermometry, and Water and Methanol Masers in IC\,342, NGC\,6946 and NGC\,2146}

\author{Mark Gorski\altaffilmark{1,2}}
\email{mgorski@unm.edu}

\author{J\"urgen Ott\altaffilmark{1,3}}
\email{jott@nrao.edu}

\author{Richard Rand\altaffilmark{2}}
\email{rjr@unm.edu}

\author{David S. Meier\altaffilmark{3,1}}
\email{david.meier@nmt.edu}

\author{Emmanuel Momjian\altaffilmark{1}}
\email{emomjian@nrao.edu}

\author{Eva Schinnerer\altaffilmark{4}}
\email{schinner@mpia.de}


\affil{1 National Radio Astronomy Observatory, P.O. Box O, 1003 Lopezville Road, Socorro, NM 87801, USA}
\affil{2 Department of Physics and Astronomy, University of New Mexico, 1919 Lomas Blvd NE, Albuquerque, NM 87131, USA}
\affil{3 Department of Physics, New Mexico Institute of Mining and Technology, 801 Leroy Place, Socorro, NM 87801, USA}
\affil{4 Max-Planck Institut f\"ur Astronomie, K\"onigstuhl 17, D-69117 Heidelberg, Germany}


 \defcitealias{Gorski2017}{G17}

\begin{abstract}

The Survey of Water and Ammonia in Nearby galaxies (SWAN) studies atomic and molecular species across the nuclei of four star forming galaxies: NGC\,253, IC\,342, NGC\,6946, and NGC\,2146. As part of this survey, we present Karl G. Jansky Very Large Array (VLA) molecular line observations of three galaxies: IC\,342, NGC\,6946 and NGC\,2146. NGC\,253 is covered in a previous paper. These galaxies were chosen to span an order of magnitude in star formation rates and to select a variety of galaxy types. We target the metastable transitions of ammonia \amm(1,1) to (5,5), the 22\,GHz water (\water) ($6_{16}-5_{23}$) transition, and the 36.1\,GHz methanol (CH$_3$OH) ($4_{-1}-3_{0}$) transition. \added{We use the \amm\ metastable lines to perform thermometry of the dense molecular gas.} We show evidence for uniform heating across the central kpc of IC\,342 with two temperature components for the molecular gas\added{, similar to NGC 253,} of 27\,K and 308\,K,  and that the dense molecular gas in NGC\,2146 \added{has} a temperature $<$86 K.  We identify two new water masers in IC\,342, and one new water maser in each of NGC\,6946 and NGC\,2146.  The two galaxies NGC\,253 and NGC\,2146, with the most vigorous star formation, host \water\ kilomasers. Lastly, we detect the first 36\,GHz \methanol\ masers in IC\,342 and NGC\,6946. For the four external galaxies the total \methanol\ luminosity in each galaxy suggests a correlation with galactic star formation rate, whereas the morphology of the emission is similar to that of HNCO, a weak shock tracer. 
\end{abstract}


\section{Introduction}

Galaxy evolution models without feedback overestimate star formation rates and efficiencies (e.g., \citealt{Kauffmann1999}). Energy and momentum injected into the interstellar medium (ISM) by stars is the dominant mechanism for impeding the formation of future generations of stars. Supernovae, stellar winds, photoionization, and shock heating are a few methods by which energy is injected into the ISM. Without these effects the ISM tends to rapidly collapse under its own gravity and form stars in less than a dynamical time, while too much feedback can completely disrupt giant molecular clouds (GMCs) (e.g., \citealt{Kauffmann1999}; \citealt{Krumholz2011}; \citealt{Hopkins2011}).  \citet{Murray2010} suggest that each of these mechanisms dominates at different times during the life of GMCs, and \citet{Hopkins2014} show that feedback mechanisms combine non-linearly so that no individual mechanism dominates. Therefore it is necessary to check the feedback prescription in cosmological simulations against observations. 

 Since star formation is largely correlated with dense molecular gas \citep{Gao2004}, the effect of feedback should, in part, be traced by the state of the dense molecular ISM. Radio interferometric observations with the VLA of nearby galaxies provide access to 10-100 pc scales related to GMCs. By using diagnostically important molecular tracers, it is possible to reveal the properties of the ISM on these relevant scales and test feedback. 

\begin{deluxetable}{lllll}
\tabletypesize{\footnotesize}
\tablewidth{0pt}
\tablecolumns{5}
\tablecaption {Adopted Galaxy Properties\label{tab:GalProp}}
\tablehead{
 Galaxy	& Distance	&  SFR$^\text{b}$	& Linear Scale 		& V$_{\rm{sys}}$ \\
 		&	(Mpc)		&	 (\sfr)		&	(pc/ \arcsec) 	& (\kms)	\\
}
\startdata
IC\,342		&	3.28$^\text{a}$ 	&	2.8	&	16	&	\phn35$^\text{d}$\\
NGC\,253		&	3.50$^\text{c}$	&	4.2	&	17	&	235$^\text{e}$\\
NGC\,6946	&	5.89$^\text{a}$ 	&	3.2	& 	29	&	\phn50$^\text{f}$\\
NGC\,2146	&	15.2$^\text{b}$	&	20	&	74	& 	850$^\text{g}$\\		
\enddata
\tablecomments{(a) \citet{Karachentsev2013}, (b) SFR is calculated from the infrared luminosity from \citet{Gao2004} where $\dot{M}_{\rm SFR} (M_\odot/yr)\approx 2\times10^{-10} (L_{\rm IR}/L_{\odot})$, e.g. \citet{Kennicutt1998}, (c) \citet{Radburn-Smith}, (d) \citet{Meier2001}, (e) \citet{W1999}, (f) \citet{Schinnerer2006}, (g) \citet{Greve2006}}
\end{deluxetable}

The  ``Survey of Water and Ammonia in Nearby galaxies" (SWAN) is a survey of molecular and atomic transitions at centimeter and millimeter wavelengths of \added{ the central $\sim$2\arcmin\ of} four galaxies:  NGC\,253, IC\,342, NGC\,6946, and NGC\,2146.   \added{The adopted properties of these galaxies are listed in Table \ref{tab:GalProp}} These galaxies were chosen to span a \replaced{range of galactic environments}{range of galactic host properties}. Here we present results for key diagnostic molecular transitions from this survey including the ammonia (\amm) metastable transitions J=K (1,1) to (5,5), the 22 GHz water (\water) ($6_{16}-5_{23}$) maser, and the 36\,GHz methanol (\methanol) ($4_{14}-3_{03}$) maser across the nuclei.

\added{ The \amm\ molecule can provide useful diagnostics of the properties of the dense molecular gas. Specifically} measurements of the \amm\ metastable transitions provide for calculations of the rotation temperature, and, via Large Velocity Gradient (LVG) models, it is possible to estimate  kinetic temperatures with reasonable precision (e.g., \citealt{H&T}, \citealt{walmsley83}, \citealt{Lebron2011}, \citealt{Ott2005}, \citealt{Ott2011}, \citealt{Mangum2013}, \citetalias{Gorski2017}). 

 The 22 GHz \water\ masers are collisionally excited, coming from three environments. The stellar masers are associated with mass loss stages of Young Stellar Objects (YSO's) and Asymptotic Giant Branch (AGB) stars and are the least luminous $<0.1$\Lsun\ (e.g., \citealt{Palagi1993}). The strongest masers ($>$20 \Lsun), classified as megamasers, are typical of nuclear activity (e.g., \citealt{Braatz1996}). They are often found in the nuclear tori of AGN (e.g., \citealt{Reid2009}). Kilomasers are of  intermediate luminosity between the stellar and megamasers, and are associated with strong star formation activity \citep{Hagiwara2001}. These could consist of many stellar class masers or be the low luminosity tail of the megamaser class \citep{Tarchi2011}.  We use these masers as signposts of shocked material related to star formation. 
 
 The Class I, collisionally excited, 36\,GHz \methanol\ masers are new in the extragalactic context. \added{Class II \methanol\ masers are radiatively pumped and are outside the scope of this paper.}
 Before this paper Class I \methanol\ masers had been detected outside the Milky Way only in NGC\,253 (\citealt{Ellingsen2014} and  \citetalias{Gorski2017}) and Arp220 \citep{Chen2015}, where they are orders of magnitude more luminous than typical galactic counterparts. \added{It is possible the masers are related to weak shocks, though much remains unknown about these masers in this context.}  
\added{NGC\,253 features strong star formation in its central kpc and is the focus of \cite{Gorski2017} (hereafter \citetalias{Gorski2017})}. In \citetalias{Gorski2017} we performed \amm\ thermometry, and analyzed \water\ and 36\,GHz \methanol\ masers, in the central kpc of NGC\,253 based on Karl G. Jansky Very Large Array (VLA)\footnote{The National Radio Astronomy Observatory is a facility of the National Science Foundation operated under cooperative agreement by Associated Universities, Inc.}  observations. We detected the \amm(1,1) to (5,5) lines, and the (9,9) line. Using the same analysis we will use in this paper we uncovered a cool 57\,K component and a warm 130\,K component uniformly distributed across the molecular disk. There is no apparent correlation of temperature with tracers of dominant forms of feedback (weak or strong shocks, and Photon Dominated Regions (PDRs)). Within the centermost 100\,pc there is evidence for \amm(3,3) masers, similar to NGC\,3079 \citep{Miyamoto2015}. The strongest water maser in NGC\,253 contains many components and shows evidence for an extension along the minor axis. This suggests a relationship with the outflow, in addition to the masers' strong blueshifted velocity components. We also resolved the first detected extragalactic 36\,GHz  masers \citep{Ellingsen2014} into five different sources. The emission is  concentrated around the edges of expanding superbubbles. The morphology of the emission is similar to HNCO($4_{0,4}-3_{0,3}$), a weak shock tracer \citep{Meier2015}, suggesting the two molecules trace similar conditions. 

In this paper we carry out a similar analysis across the other galaxies in the SWAN sample (IC\,342, NGC\,6946, and NGC\,2146). \added{The spiral galaxy IC\,342 has a relatively modest total global star formation rate of 2.8 \sfr. The central kpc has a molecular mass of $\sim4\times10^7$\,M$_\odot$ and consists of two molecular spiral arms that terminate in a central molecular ring (e.g., \citealt{Downes1992}, \citealt{Turner1992}, and \citealt{Meier2005}).

NGC\,6946 is a nearby spiral galaxy with a star formation rate of 3.2 \sfr,  a nucleated starburst that is being fed by inflows along a molecular bar, and a molecular mass of $\sim3.1\times10^8$\,M$_\odot$\ within the central kpc \citep{Schinnerer2006}. 
 
 Lastly NGC\,2146 is a nearby starburst with a star formation rate of 20 \sfr. It is peculiarly warped with no obvious companion, has a large reservoir of molecular gas ($\sim4.1\times10^9\,\text{M}_\odot$), and a molecular outflow \citep{Tsai2009}. The adopted properties of these galaxies are listed in Table \ref{tab:GalProp}. }

In \S 2 we describe the observational setups, and the data reduction and imaging process. In \S 3 we report our measurements of the \amm, \water, and \methanol\ lines for each galaxy individually. In \S 4 we discuss the analysis of the \amm\ lines and derived temperatures for each galaxy where possible; in addition we generate LVG models for IC\,342. We also discuss the \water\ and \methanol\ masers in each galaxy.  In \S 5 we incorporate the results from \citetalias{Gorski2017} and discuss the relevance of the survey as a whole. Finally, in \S 6 we summarize our findings.


\section{Observations and Data Reduction}

 IC\,342 and NGC\,6946 were observed with identical frequency setups of the VLA (project code: 10B-161). The correlator was set up to cover frequency ranges 21.9$-$22.3 GHz, 23.6$-$23.7 GHz, and 24.1$-$24.5 GHz with 500\,kHz wide channels (K-band receiver), and 26.8$-$27.9 GHz and 36.0$-$37.0 GHz with 1\,MHz  wide channels (Ka-band receiver). The galaxy NGC\,6946 was observed with two pointings with the same RA(J2000): 20$^h$\,35$^m$\,52.336$^s$ and different DEC(J2000): +60\arcdeg\,09'\,32.2100'' and +60\arcdeg\,08'\,56.210''. The galaxy NGC\,2146 was  observed with the same correlator setup as NGC\,253 in \citetalias{Gorski2017} (project code: 13A-375), with 250\,kHz channels, but was not observed in Ka-band. All three galaxies were observed in the C configuration of the VLA. \added{The C configuration has a minimum baseline length of 0.35 km. At 22\,GHz and 36\,GHz this means scales above 66\arcsec\ and 44\arcsec\ are respectively resolved out.} As in \citetalias{Gorski2017} we target the metastable \amm\ lines from J=K (1,1) to (5,5), the 22.2351\,GHz \water($6_{16}-5_{23}$)  maser, and the 36.1693\,GHz \methanol($4_{14}-3_{03}$) maser. NGC\,2146 did not get observed in Ka-band, so no analysis of the 36\,GHz \methanol\ maser is presented for this galaxy. \added{The rest frequencies for our targeted molecular transitions are shown in Table \ref{tab:restfreq}.}

We calibrated and imaged the data in the Common Astronomy Software Applications (CASA) package version 4.6.0 \citep{mcmullin07}. For all three galaxies J0319+4130 was observed as a bandpass calibrator. For the complex gain and flux density calibrators we used J0304+6821 and 3C147 for IC\,342, J2022+6136 and 3C48 for NGC\,6946, and J6726+7911 and 3C147 for NGC\,2146. Continuum subtraction was performed in the $(u,v)$ domain by selecting line free channels in each sub-band to define the continuum model.  Image cubes were made with natural weighting, CLEANed to $\sim3\sigma$ rms noise, primary beam corrected, and  smoothed to common spatial and spectral resolutions per galaxy. All velocities in this paper will be in the LSRK frame unless otherwise stated.  \added{ The properties of the generated data cubes are listed in Table \ref{tab:imgcube}}. \deleted{For IC\,342  the we smooth to round synthesized beams of 1.2\arcsec\ (19\,pc) for \water\ and \amm, and 1.0\arcsec\ (16\,pc) for \methanol. For NGC\,6946 the beams are 1.5\arcsec\ (43\,pc) for \water\ and \amm, and 1.0\arcsec\ (29\,pc) for \methanol. These two galaxies were smoothed to a velocity resolution of 7\kms. For IC\,342, the rms noise in the image cubes is 0.45\,mJy beam$^{-1}$ channel$^{-1}$ for \methanol\ and 0.31\,mJy beam$^{-1}$ channel$^{-1}$ for \water\ and \amm. In the  NGC\,6946 image cubes the rms noise is 0.42\,mJy beam$^{-1}$ channel$^{-1}$ for \methanol\ and 0.30\,mJy beam$^{-1}$ channel$^{-1}$ for \water\ and \amm. All image cubes for NGC\,2146 were smoothed to a round 1.5\arcsec\ (111\,pc) beam and 4\kms\ resolution, and the rms noise is 0.7\,mJy beam$^{-1}$ channel$^{-1}$. Peak flux maps were made for all three galaxies in all the lines of interest. For a 1\arcsec\ beam at 22 GHz, 1 mJy beam$^{-1}$ is 3 mK. }

\begin{deluxetable}{lc}
\tablecaption{ Molecular  Transitions\label{tab:restfreq}}
\tabletypesize{\scriptsize}
\tablewidth{0pt}
\tablehead{
\colhead{Transition} & \colhead{Rest Frequency} \\
	& (GHz)  
}
\startdata
{\water($6_{16}-5_{23}$) }			& {22.2351} 	\\
{\amm\,(1,1)} 		& {23.6945} 	\\
{\amm\,(2,2)}		&{23.7226}	\\
{\amm\,(3,3)}		& {23.8701}		\\
{\amm\,(4,4)} 		& {24.1394} 	\\
{\amm\,(5,5) }		& {24.5330}	\\
{\methanol\,($4_{-1}-3_{0}$)}	&	{36.1693}	\\
\enddata
\end{deluxetable}

\begin{deluxetable*}{lccc}
\tabletypesize{\footnotesize}
\tablewidth{0pt}
\tablecolumns{4}
\tablecaption{ Image Cube Properties.\label{tab:imgcube}}
\tablehead{
\colhead{Molecule} & \colhead{Spatial Resolution} & \colhead{Spectral Resolution} & \colhead{RMS noise} \\
	&	\arcsec	& \kms\ & mJy beam$^{-1}$ channel$^{-1}$ 
}
\startdata
\cutinhead{IC 342} 
\water\ \& \amm	&	1.2		&	7	& 0.31 	\\	
\methanol	&	1.0		&	7	& 0.45 	\\
\cutinhead{NGC\,6946}
\water\ \& \amm	&	1.5		&	7	& 0.30 	\\	
\methanol	&	1.0		&	7	& 0.42 	\\
\cutinhead{NGC\,2146}
\water\ \& \amm	&	1.5		&	4	& 0.70 	\\	
\enddata
\tablecomments{For a 1\arcsec\ beam at 22 GHz, 1 mJy beam$^{-1}$ is 2.5 K. The properties of the detected lines are shown in Tables \ref{tab:342amm}, \ref{tab:342water}, and \ref{tab:342methanol}.}
\end{deluxetable*}

\section{Results}

We report the results from each galaxy separately in the sections below. Our analysis will focus only on \amm, \water, and \methanol\ lines \added{(Table \ref{tab:restfreq})}. We identified the emission by peaks in the peak flux maps. \added{The resolution for each map and rms for the parent data cube are shown in Table \ref{tab:imgcube}.} We have chosen a conservative detection threshold with an integrated flux of \replaced{20 mK \kms}{20 K \kms}\explain{wrong units}. \amm\ emission was primarily identified using the \amm(3,3) peak flux maps as \amm(3,3) is usually the strongest line. We also search the \amm(1,1) peak flux map for emission, because being the lowest energy transition it is the easiest to excite. \added{ The locations are listed in Tables \ref{tab:342amm}, \ref{tab:342water}, and \ref{tab:342methanol}.}\added{ We fit gaussian profiles to the emission lines where appropriate. Note that some spectral lines are narrow resulting greater uncertainties than the better sampled broader spectral lines. In general the measured line width is deconvolved from the spectral resolution. }

\subsection{IC\,342}

\begin{figure*}
\includegraphics[width=0.95\textwidth]{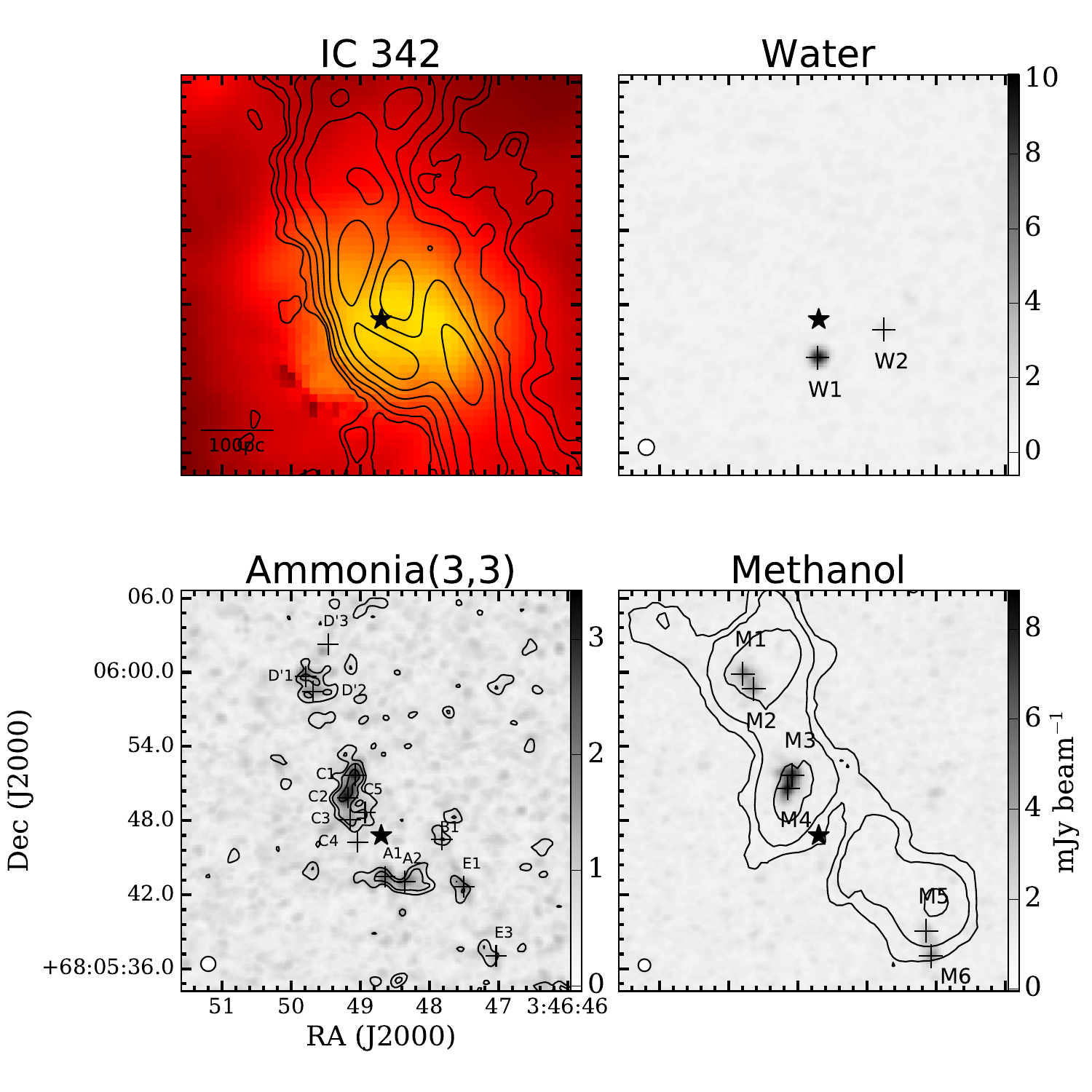}
\caption{Here we show the centermost 16\arcsec of IC\,342.  At 23 GHz the diameter of the primary beam of the VLA is 1.9\arcmin.  We are showing $\sim$10\% of the total field of view of the VLA where detections are made. Spitzer IRAC 8$\mu$m Image of IC\,342 from the Infrared Science Archive (color) and \aco\ 3, 5, 9, 15, 24 and 39 $\times$ 10 Jy beam$^{-1}$ \kms\ contours from \citet{Meier2000} (top left), \water\  peak flux map (top right), \amm(3,3)  peak flux map with \amm(3,3) 3, 5, and 9 $\times$ 9 mJy beam$^{-1}$ \kms\ contours (bottom left), and \methanol\ peak flux map with HNCO($4_{04}-3_{03}$) 3, 5, and 9 $\times$ 0.15 Jy beam$^{-1}$ \kms\ contours from \citet{Meier2000} (bottom right). The star shows the dynamical center ($\alpha_{2000}$ = 03:46:48.7, $\delta_{2000}$ = +68 05 46.8; \citealp{Turner1992}). Peaks where spectra were extracted are labeled with plus signs.}
\label{fig:342select}
\end{figure*}

\begin{figure*}
\includegraphics[width=0.999\textwidth]{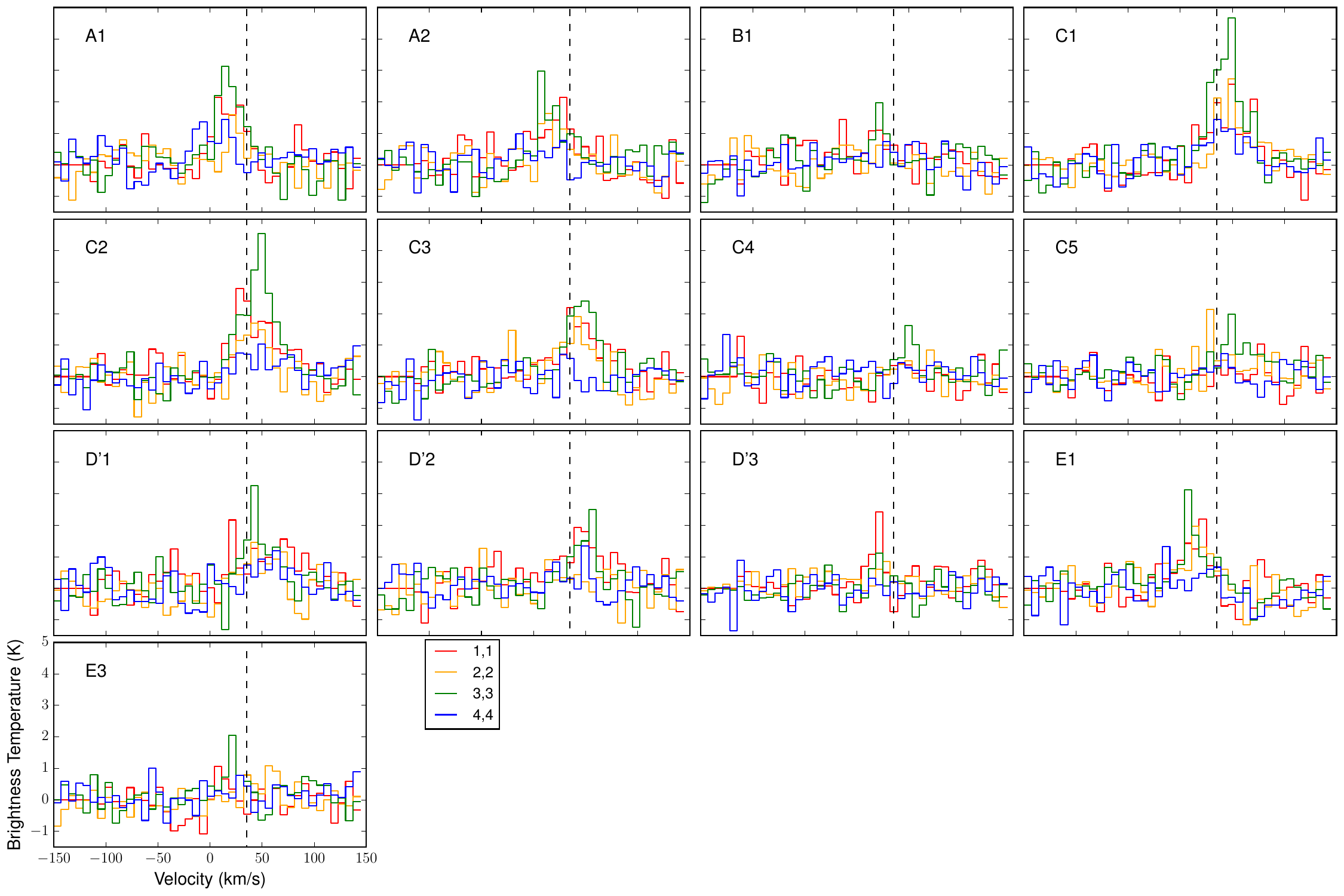}
\caption{The \amm(1,1) to (4,4) spectra extracted from the peaks in IC\,342 marked in Figure \ref{fig:342select}. The vertical dashed line shows the systemic velocity of 35 \kms. }
\label{fig:342_amm_spec}
\end{figure*}

The central kpc of IC\,342 consists of two molecular spiral arms that terminate in a central molecular ring (Figure \ref{fig:342select}).  Six main molecular clouds have been identified and discussed in the linterature:  A, B, C, D, D$'$, and E  (e.g., \citealp{Downes1992}, \citealp{Meier2000}, \citealp{Meier2005}), and these are all detected in \amm\ in our observations. The northern arm consists of the C, D, and D$'$ clouds, spanning velocities from ~44\kms\ to 62\kms, and the southern arm consists of the A, B, and E clouds, spanning velocities from ~13\kms\ to 27\kms. In some clouds we have observed substructure resulting in multiple peaks in the \amm(3,3) peak flux map (Figure \ref{fig:342select}). The \amm\ spectra are shown in Figure \ref{fig:342_amm_spec}. We utilize the \amm(3,3) line for identification because it is often the strongest transition. In cases where we do not detect \amm(3,3) we use the \amm(1,1) line. We do not detect the \amm(5,5) line and not all metastable transitions J=K $<$ 4 are detected in all locations. We have extracted spectra from each of these peaks and fit single Gaussians to the spectra. We do not detect the hyperfine \amm\ lines. We adopt the naming convention from \cite{Meier2001}. We do not detect the E2 region from \citet{Meier2011} in \amm, however we detect a few  new peaks in the A, C, D$'$, and E clouds.   Their positions and extracted integrated flux, line center, FWHM, and peak flux are tabulated in Table \ref{tab:342amm}.

\begin{figure}
\includegraphics[width=0.5\textwidth]{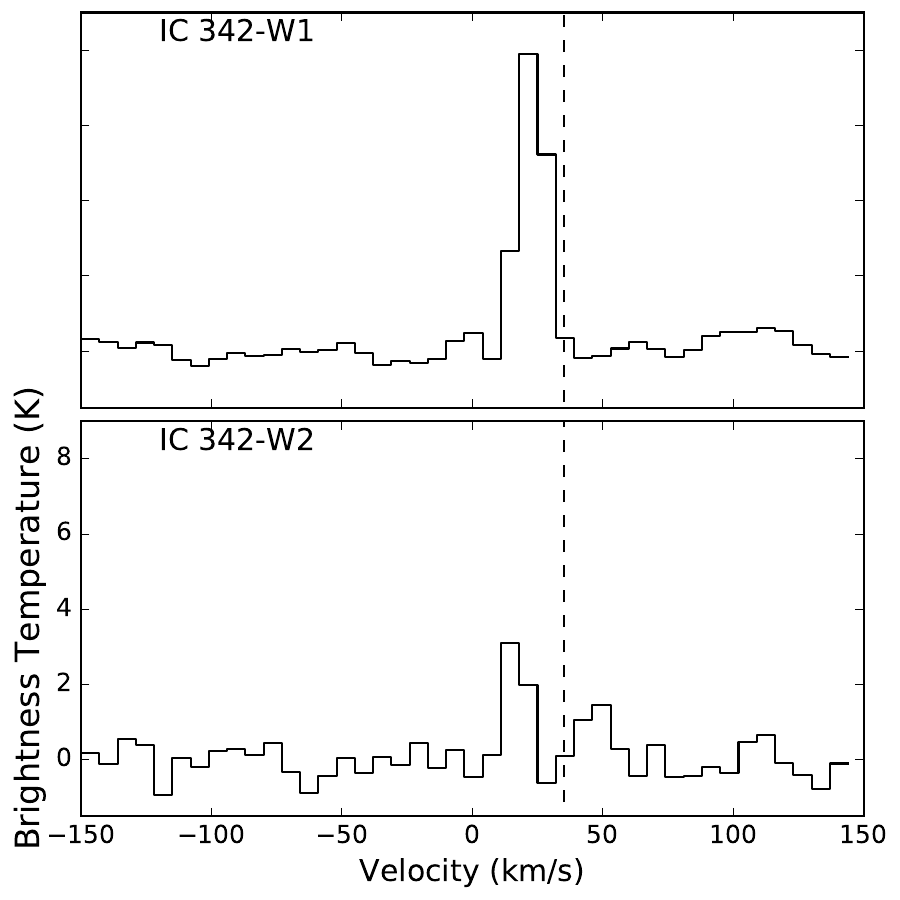}
\caption{The 22 GHz \water\ maser spectra extracted from the peaks in IC\,342 marked in Figure \ref{fig:342select}. The vertical dashed line shows the systemic velocity of 35 \kms. }
\label{fig:342_water_spec}
\end{figure}

We have identified two water masers in IC\,342: IC\,342-W1 in the A cloud and IC\,342-W2 in the B cloud (Figure \ref{fig:342select}). The spectra are shown in Figure \ref{fig:342_water_spec}. Water masers in IC\,342 have already been investigated by \citet{Tarchi2002}. They found a single narrow water maser to the east of the B cloud ($\alpha_\text{2000}$ = 03$^h$46$^m$46.3$^s$, $\delta_\text{2000}$ =+68\arcdeg05\arcmin46\arcsec, with a positional uncertainty of $\sim 5$\arcsec). We do not detect this maser. It is likely variable as they observed a peak flux increase of $\sim$100\% over 20 days. Our masers have narrow spectral profiles, being almost single channel (7\kms) detections, and have isotropic luminosities of $<$\,0.02\Lsun. The spectrum of W1 was Hanning smoothed to reduce Gibbs ringing. The properties of the spectral profiles are listed in Table \ref{tab:342water}.

\begin{figure*}
\includegraphics[width=0.99\textwidth]{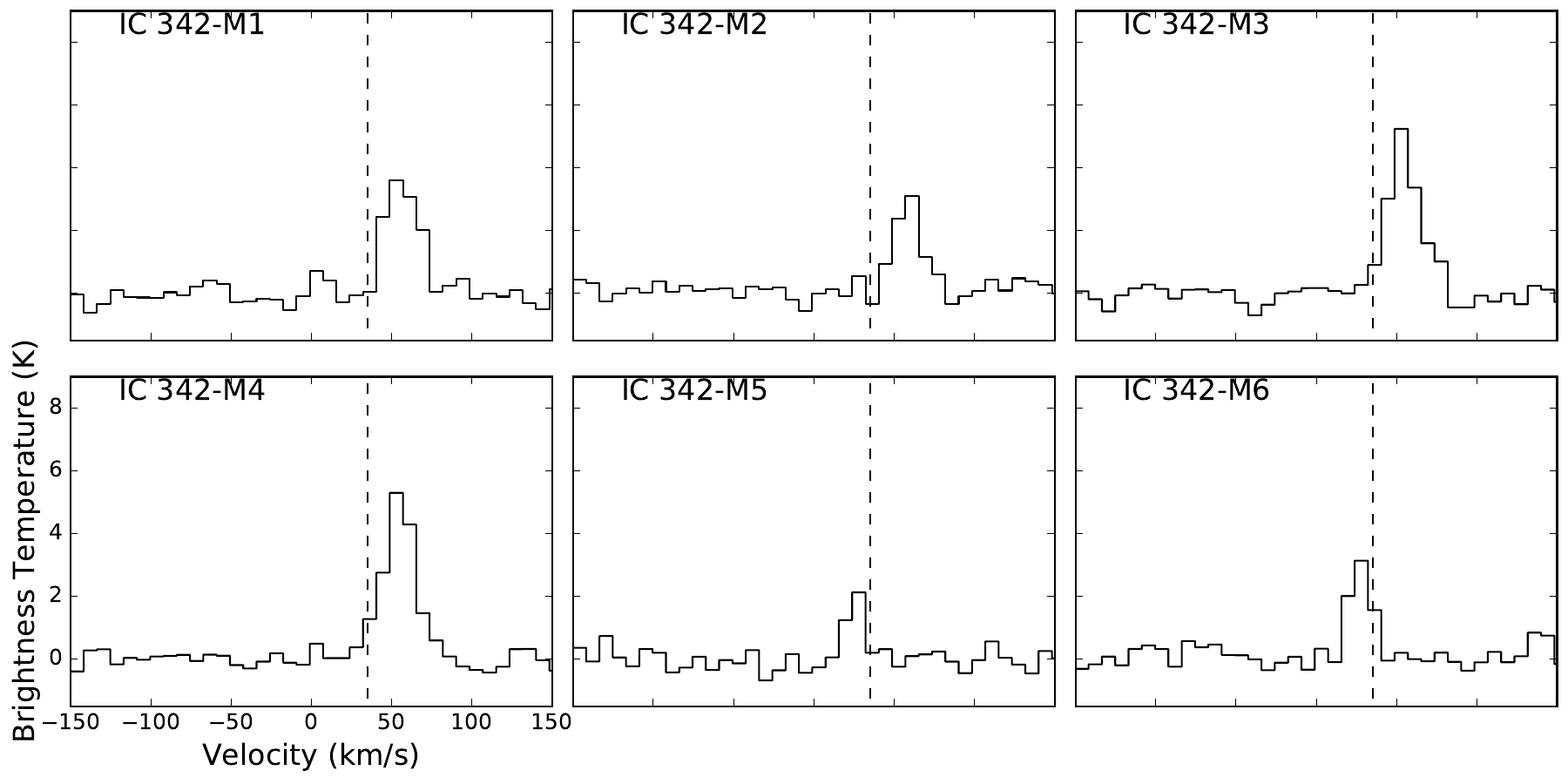}
\caption{The 36 GHz \methanol\  spectra extracted from the peaks in IC\,342 marked in Figure \ref{fig:342select}. The vertical dashed line shows the systemic velocity of 35 \kms. }
\label{fig:342_methanol_spec}
\end{figure*}

We now add IC\,342 to the list of galaxies with detected 36\,GHz \methanol\ emission. We detect six spatially resolved 36\,GHz \methanol\ sites in IC\,342. There are two in each of the D, C, and E clouds (Figure \ref{fig:342select}). The spectra are shown in Figure \ref{fig:342_methanol_spec} and the extracted properties are shown in Table \ref{tab:342methanol}. These masers have isotropic luminosities of order 10$^{-2}$ \Lsun, about 10-100 times less luminous than the masers in NGC\,253 \citepalias{Gorski2017}.

\subsection{NGC\,6946}

\begin{figure*}
\includegraphics[width=0.9\textwidth]{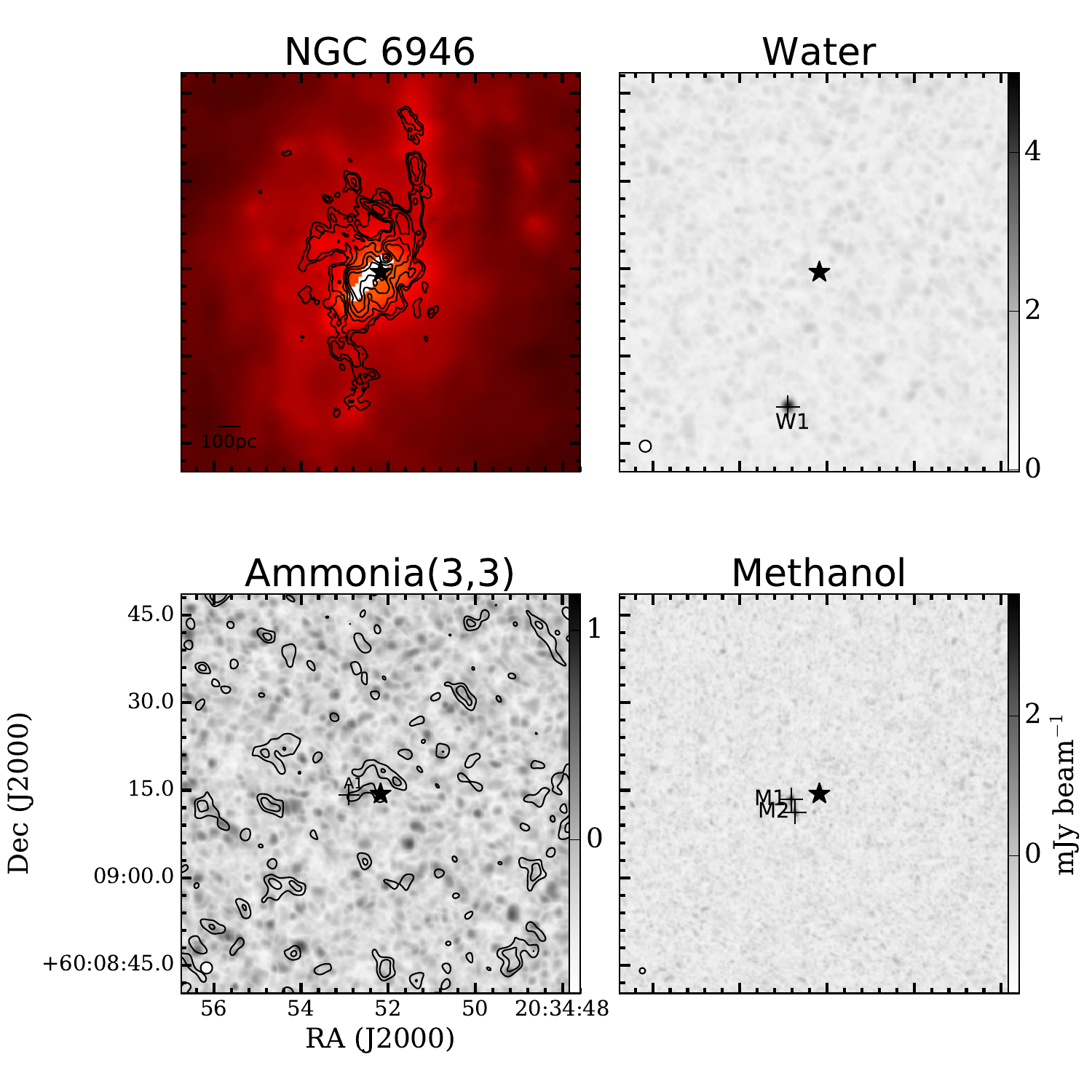}
\caption{Here we show the centermost 34\arcsec of IC\,342.  At 23 GHz the diameter of the primary beam of the VLA is 1.9\arcmin.  We are showing $\sim$30\% of the total field of view of the VLA where detections are made. Spitzer IRAC 8$\mu$m  image of NGC\,6946  from the Infrared Science Archive (color) and \aco\ 3, 5, 9, 15, 24 and 39 $\times$ 0.05 Jy beam$^{-1}$ \kms\ contours from \citet{Schinnerer2006} (top left), \water\ peak flux map (top right), \amm(3,3) peak flux map (greyscale) with \amm(3,3) integrated flux contours at 0.03 and 0.06 Jy \kms\  (bottom left), and \methanol\ (bottom right) peak flux maps. The star shows the peak of the 22\,GHz continuum. Peaks where spectra were extracted are labeled with plus signs. }
\label{fig:6946select}
\end{figure*}

\begin{figure}
\includegraphics[width=0.5\textwidth]{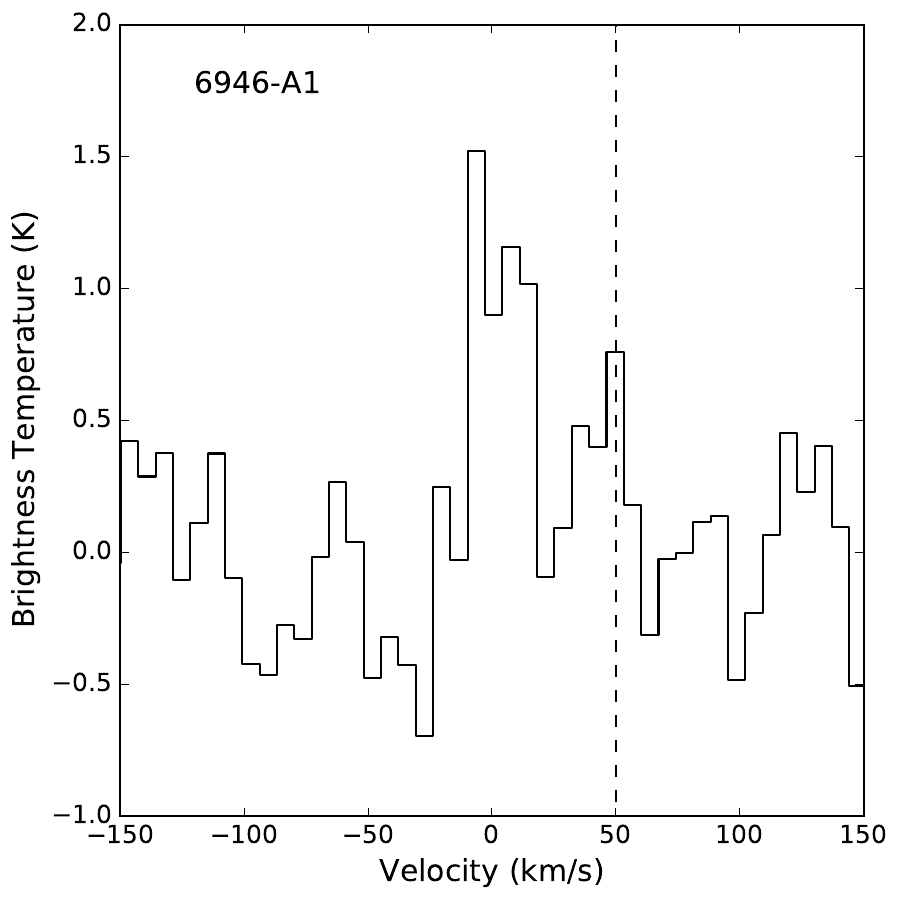}
\caption{The \amm(3,3) spectrum extracted from the peak in NGC\,6946 marked in Figure \ref{fig:6946select}. The vertical dashed line shows the systemic velocity of 50 \kms. }
\label{fig:6946_amm_spec}
\end{figure}

The molecular gas in the central kpc of NGC\,6946 (Figure \ref{fig:6946select}) shows spiral structure with a nuclear bar that feeds the nuclear starburst, and three clumps near the dynamical center \citep{Schinnerer2006}. These structures are not detected in our \amm\ observations. Our only detection is the \amm(3,3) line $\sim$4.5\arcsec\ from the dynamical center (Figure \ref{fig:6946select}). The integrated line intensity is 21.8$\pm$3.8 K \kms, just above our specified detection threshold. 
The line properties are listed in Table \ref{tab:342amm} and the spectrum is shown in Figure \ref{fig:6946_amm_spec}.
\begin{figure}
\includegraphics[width=0.5\textwidth]{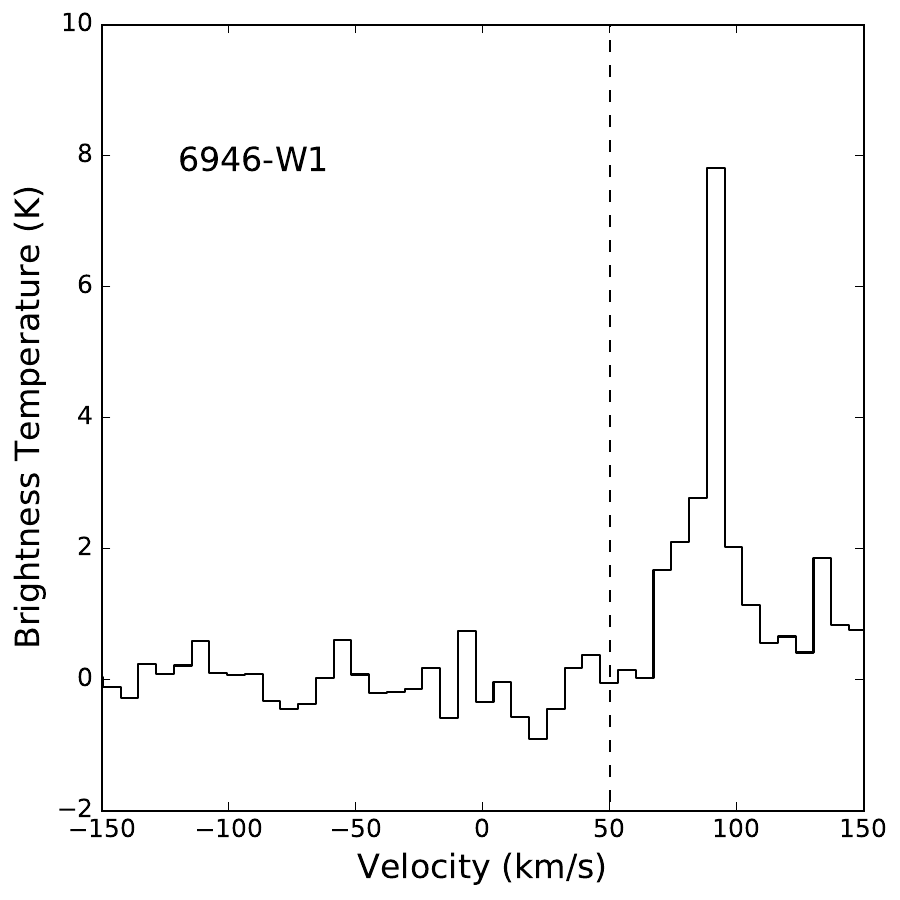}
\caption{The 22 GHz \water\ maser spectrum extracted from the peak in NGC 6946 marked in Figure \ref{fig:6946select}. The vertical dashed line shows the systemic velocity of 50 \kms. }
\label{fig:6946_water_spec}
\end{figure}

We observe a single water maser in NGC\,6946. The spectrum is narrow with a fitted FWHM of 11\kms\ and an isotropic luminosity of 0.042 \Lsun\ (Figure \ref{fig:6946_water_spec}). It is located $\sim$22\arcsec\ from the center of the galaxy in the southern spiral arms (Figure \ref{fig:6946select}). Its spectral properties are collated in Table \ref{tab:342water}. 

\begin{figure}
\includegraphics[width=0.5\textwidth]{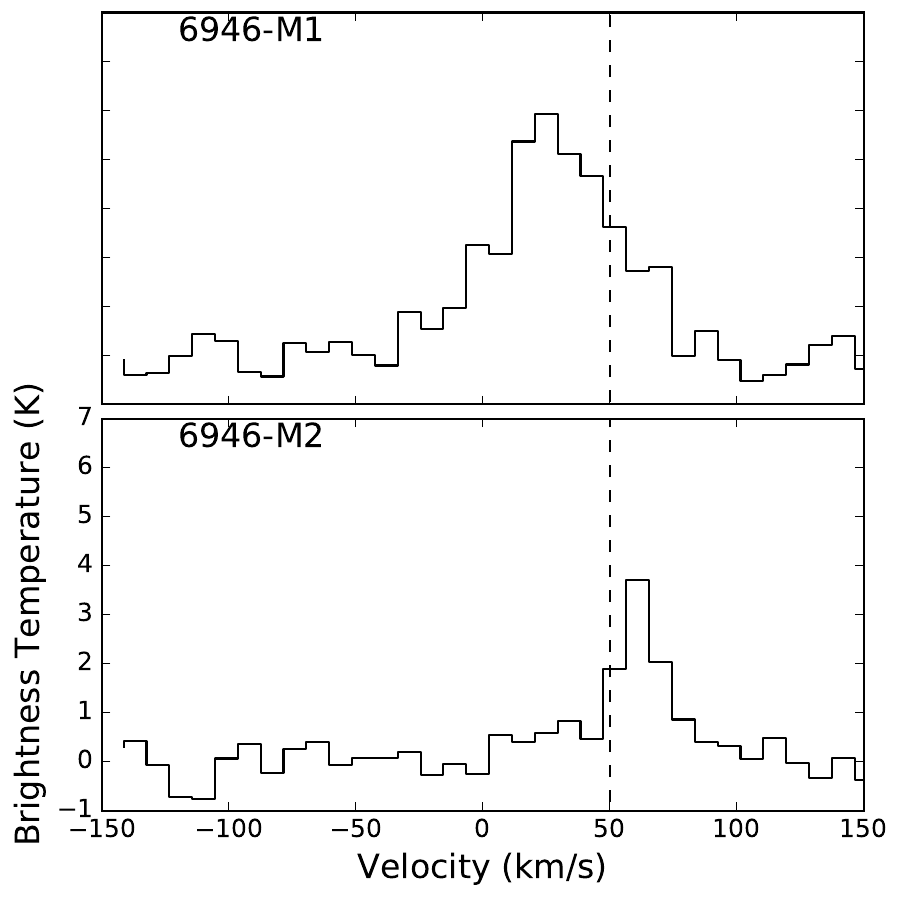}
\caption{The 36 GHz \methanol\ maser spectrum extracted from the peak in NGC\,6946 marked in Figure \ref{fig:6946select}. The vertical dashed line shows the systemic velocity of 50 \kms. }
\label{fig:6946_methanol_spec}
\end{figure}

We also expand the detections of extragalactic 36\,GHz \methanol\ masers to include NGC 6946. There are two masers located in the southern clump identified by \citet{Schinnerer2006}. Their spectral properties are available in Table \ref{tab:342methanol} and the spectra are shown in Figure \ref{fig:6946_methanol_spec}. They are one order of magnitude more luminous than the masers in IC\,342, more comparable to the ones found in NGC\,253. 

\subsection{NGC\,2146}

\begin{figure*}
\includegraphics[width=0.9\textwidth]{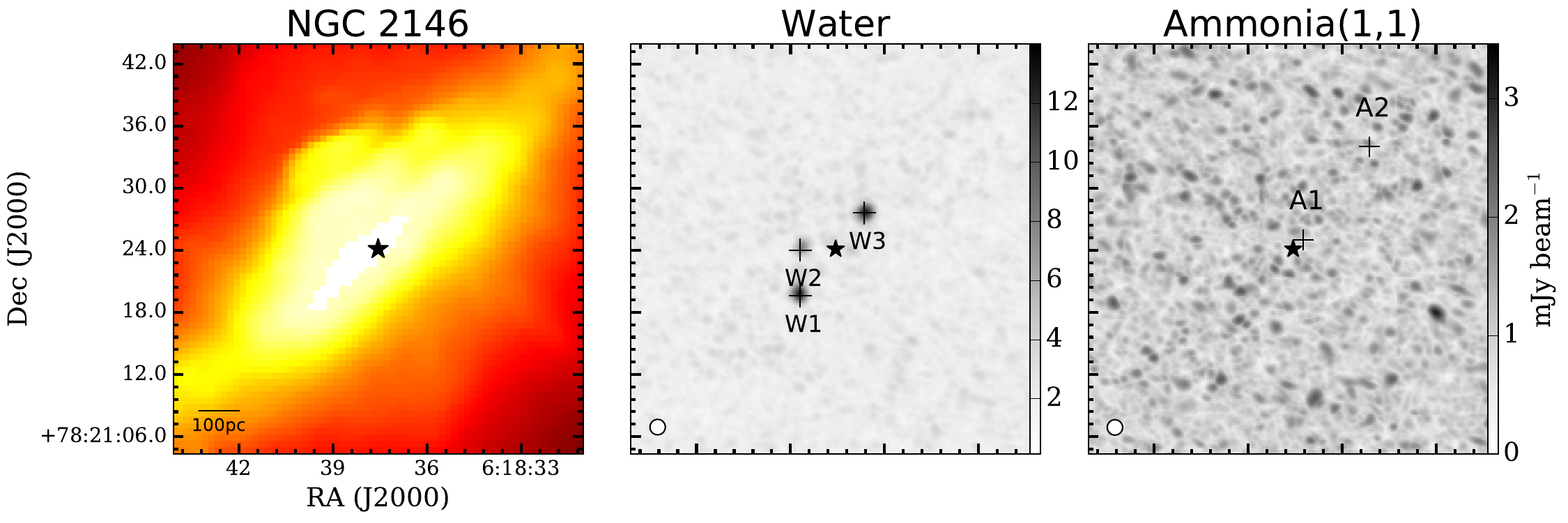}
\caption{Here we show the centermost 20\arcsec of IC\,342.  At 23 GHz the diameter of the primary beam of the VLA is 1.9\arcmin.  We are showing $\sim$17\% of the total field of view of the VLA where detections are made.NGC\,2146 image of the Spitzer IRAC 8$\mu$m emission (color) from the Infrared Science Archive (left), and \water\ (center) and \amm(3,3) (right) peak flux maps. The star shows the peak of the 22\,GHz continuum. Peaks where spectra were extracted are labeled with plus signs.}
\label{fig:2146select}
\end{figure*}

\begin{figure}
\includegraphics[width=0.5\textwidth]{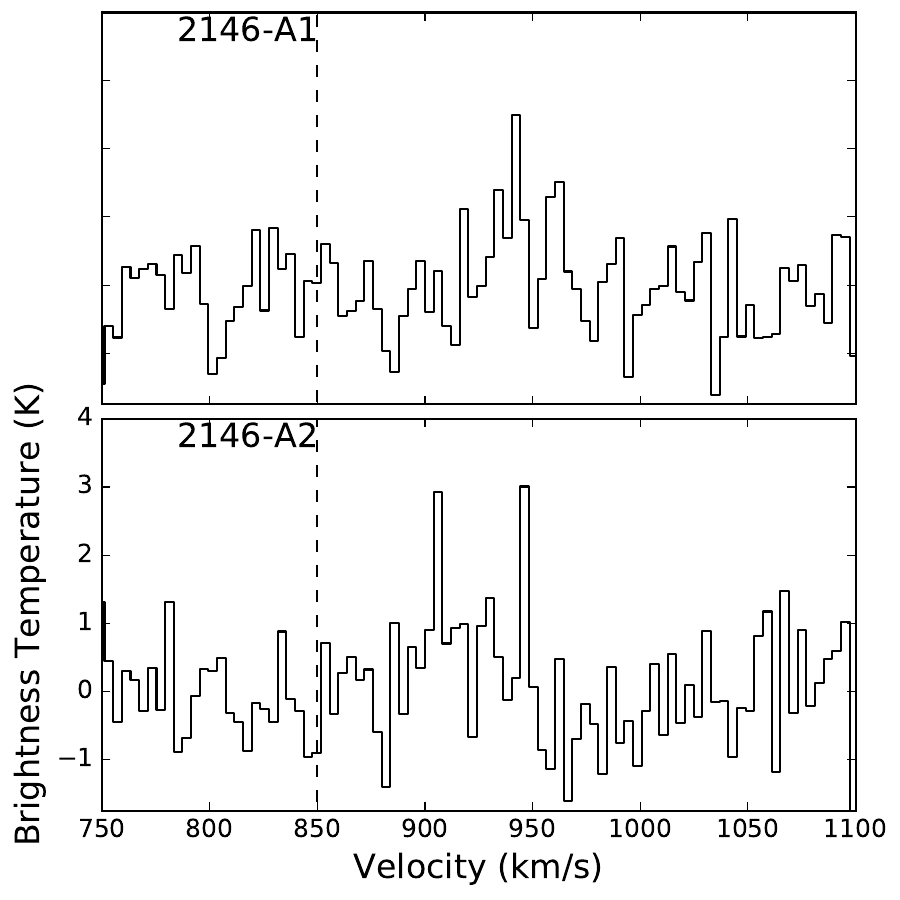}
\caption{The \amm(1,1) spectra extracted from the peaks in NGC\,2146 marked in Figure \ref{fig:2146select}. The vertical dashed line shows the systemic velocity of 850 \kms. }
\label{fig:2146_amm_spec}
\end{figure}

This peculiar spiral galaxy (Figure \ref{fig:2146select}) has the strongest star formation activity in our survey (20 \sfr; \citet{Gao2004}).  We detect the \amm(1,1) line in two locations about the center of the galaxy (tabulated in Table \ref{tab:342amm}). The spectra are shown in Figure \ref{fig:2146_amm_spec}. The peaks (Figure \ref{fig:2146select}) do not coincide with the molecular outflow or superbubbles found by \cite{Tsai2009}.

\begin{figure}
\centering
\includegraphics[width=0.5\textwidth]{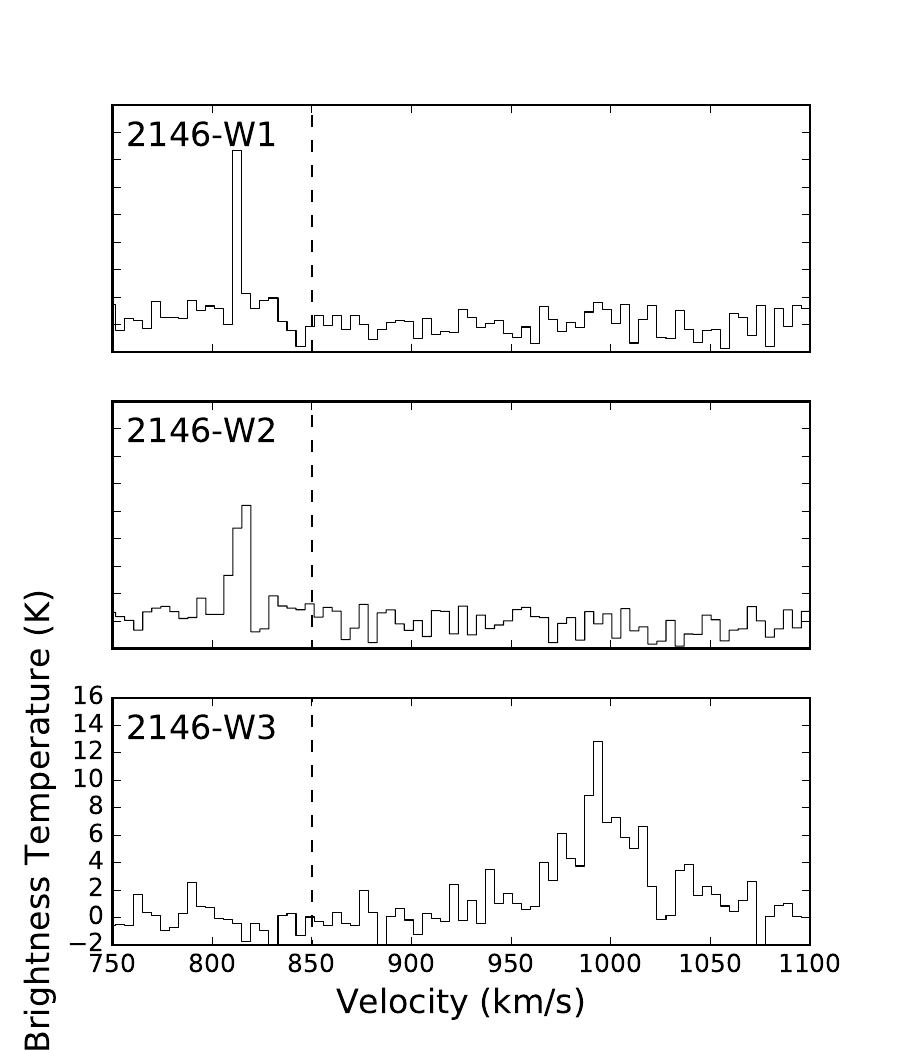}
\caption{The 22\,GHz \water\ \added{maser} spectra extracted from the peaks in NGC\,2146 marked in Figure \ref{fig:2146select}. The vertical dashed line shows the systemic velocity of 850 \kms. }
\label{fig:2146_water_spec}
\end{figure}

 There are three distinct locations with \water\ masers, all within 7\arcsec\ from the  22\,GHz continuum peak (Figures \ref{fig:2146select} and \ref{fig:2146_water_spec}). Masers 2146-W1 and 2146-W2 are narrow, almost single channel ($\sim$10\kms\ FWHM), detections whereas 2146-W3 has a broad profile with a FWHM of 48\kms. The \water\ masers in NGC\,2146 are more luminous than those in the other galaxies. Each maser's luminosity is $>$0.1\Lsun, classifying them as kilomasers.  2146-W3 is the most luminous at $\sim$2.1 \Lsun. The fitted line profiles are listed in Table \ref{tab:342water}. 2146-W2 and 2146-W3 were observed by \cite{Tarchi2002b}  (labeled respectively 2146-A and 2146-B; \citealp{Tarchi2002b} ) and are spatially coincident with optically thick \HII\ regions.  2146-W1 has not been previously detected. The dense, warped molecular ring within 2 kpc of the center has rotation velocities spanning $\pm$250\,\kms\ about the systemic velocity of 850\,\kms\ \citep{Tsai2009}. W1 has a velocity of $\sim$831\,\kms, W2 has a velocity of $\sim$832\,\kms, and W3 has a velocity of $\sim$1013\,\kms. All three masers have velocities consistent with the molecular ring.


\section{Discussion}
	

\subsection{Temperature Determination of the Dense Molecular Gas}

The metastable \amm\ lines are close in frequency with less than 5\% difference between the (1,1) and (5,5) lines. This means that the relative amount of resolved out flux among states is negligible if they trace the same gas, and that we probe similar sensitivities and spatial scales. If the gas is optically thin, a column density of the upper state can be calculated from the main beam brightness temperature (T$_{\rm mb}$), frequency in GHz ($\nu$), and J,K angular momentum quantum numbers:
\begin{equation}
N_{\rm u}(J,K)=\frac{ 7.73\times10^{13}}{\nu} \frac{J(J+1)}{K^2} \int T_{\rm mb}\,dv
\end{equation}
Rotational temperatures can then be determined from any pair of metastable transitions J and J$^\prime$ assuming differences in excitation temperatures between states are negligible  (e.g., \citealp{Henkel2000}, \citealp{Ott2005}, \citealp{Mangum2013} and \citetalias{Gorski2017}): 
\begin{equation}
\frac{N_{\rm u}(J^\prime,J^\prime)}{N_{\rm u}(J,J)} = \frac{g^\prime_{\rm op} (2J^\prime+1)} {g_{\rm op} (2J+1)} \exp\Big( \frac{-\Delta E}{T_{JJ^\prime}}\Big)
\label{eq:ratio}
\end{equation}
Where $N_{\rm u}$(J,J) is the column density of the upper inversion state of the (J,J) transition in cm$^{-2}$, $\Delta E$ is the difference in energy between J and J$^\prime$ states in Kelvin, and $g_{op}$ is the statistical weight of the \amm\, species. $g_{\rm op}$ = 1 for para-\amm\, where J$\neq3$n (where n is an integer),  and $g_{\rm op}$ = 2 for ortho-\amm, where J=3n with the (0,0) state belonging to ortho-\amm.
\begin{figure}[!h]
\centering
\includegraphics[width=0.5\textwidth]{342Boltz.pdf}
\includegraphics[width=0.5\textwidth]{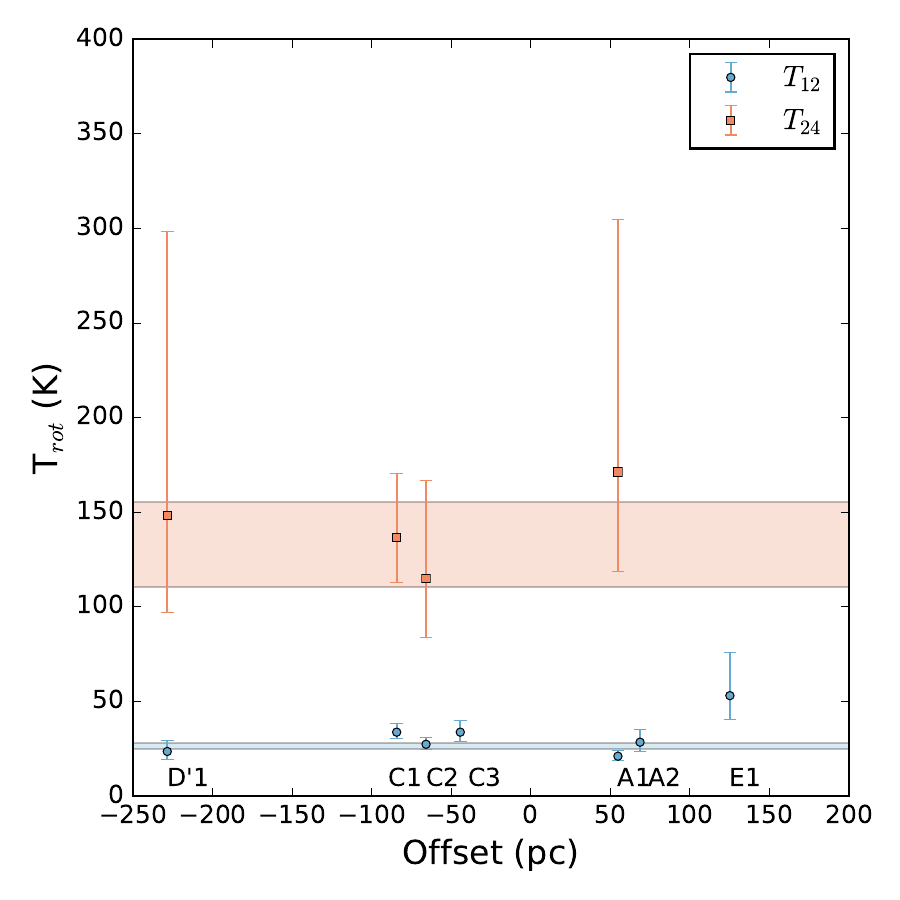}
\caption{Top: the Boltzmann diagram generated from the IC\,342 \amm\ data. The average error bars from all measurements are plotted in black along the bottom. Bottom: rotational temperatures derived from the slopes of the Boltzmann diagram. The horizontal bars represent the best fit rotational temperature across all locations. The height of the region shows the 1$\sigma$ uncertainty. }
\label{fig:342boltz}
\end{figure}

A Boltzmann diagram plots the  log of the weighted column densities on the y-axis versus the energy above ground state in Kelvin on the x-axis. The slopes between transitions then represent the inverse of the rotational temperature of the gas with colder gas represented by steeper slopes (Equation \ref{eq:ratio}). \added{Because it is a different species of \amm, We do not use \amm(3,3) for temperature determinations.} IC\,342 is the only galaxy in our sample where we have enough \amm\ detections to construct a Boltzmann diagram, shown in Figure \ref{fig:342boltz}.  
We do not construct Boltzmann diagrams for NGC\,6946 and NGC\,2146.  In NGC\,2146 we detect  only the \amm(1,1) transition. Toward NGC\,6946 we detect only the \amm(3,3) line near the center.

The rotational temperature serves as a lower limit to the true, i.e., kinetic temperature, of the gas. To estimate the kinetic temperature, we apply the approximation to Large Velocity Gradient models (LVG) by \citet{Ott2005} for the (1,1) and (2,2)  line ratio, and from \citetalias{Gorski2017} for the (2,2) and (4,4) line ratio.  \added{The LVG approximation assumes a large velocity gradient such that there is no self-absorption.}  In cases where we have only one measured transition we provide an upper limit (Table \ref{tab:342temp}). 

\begin{figure}[h]
\centering
\includegraphics[width=0.5\textwidth]{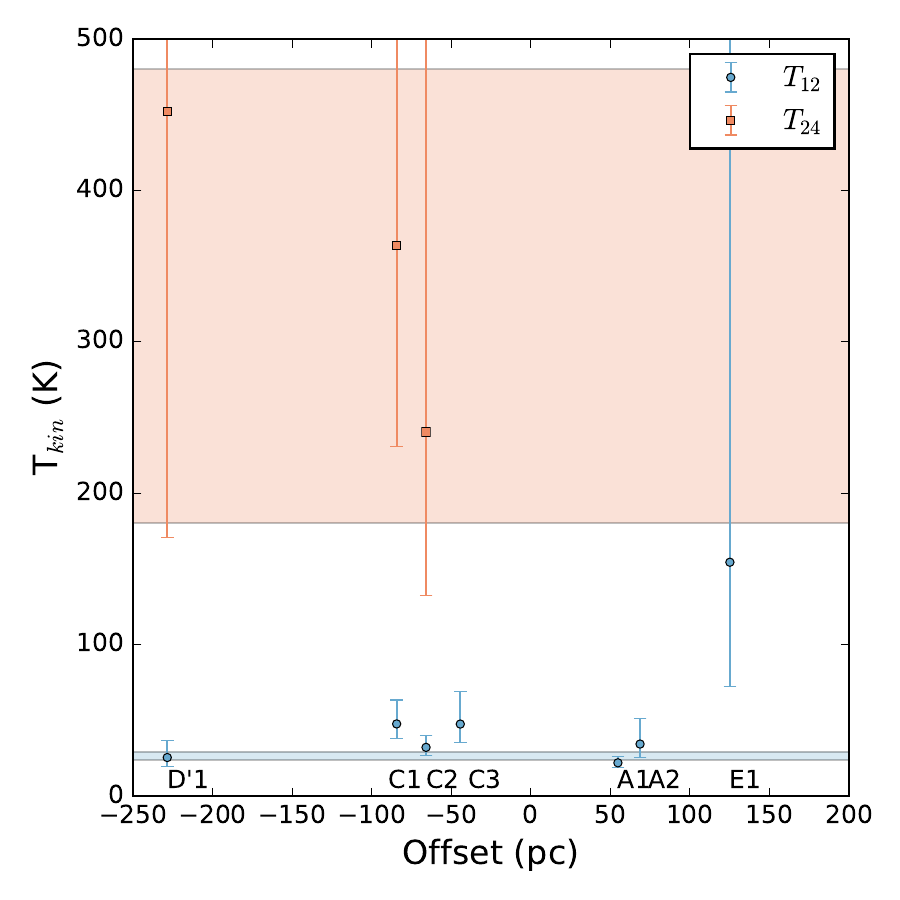}
\caption{Kinetic temperatures estimated by applying the approximation to the LVG model. The horizontal regions represent the best fit temperature to the line pairs; the height represents the 1$\sigma$ uncertainty. There is clear need for a two temperature fit with T$_{\rm kin12}$=27$\pm$3 K and T$_{\rm kin24}$=308$\pm$171 K.}
\label{fig:342TvDplot}
\end{figure}
Figure \ref{fig:342boltz} plots the rotational temperatures versus the distance from the dynamical center  ($\alpha_{2000}$ = 03:46:48.7, $\delta_{2000}$ = +68 05 46.8; \citealp{Turner1992}). \replaced{We perform a single temperature fit to each line pair weighted by the larger of the 1$\sigma$ asymmetric uncertainties}{We fit a single temperature across all locations weighted by the larger of the 1$\sigma$ asymmetric uncertainties from the individual temperatures}. The blue and red shaded areas represent the average  best fit temperatures weighted by the larger of the asymmetric errors across all detected regions from the (1,1) to (2,2) and (2,2) to (4,4) ratios, respectively. In IC\,342 the best fit rotational temperatures are T$_{12}$=26$\pm$2 K and T$_{24}$=132$\pm$22 K.  Following the method from \citetalias{Gorski2017} we convert to kinetic temperatures, finding T$_{\rm kin12}$=27$\pm$3 K and T$_{\rm kin24}$=308$\pm$171 K. T$_{\rm kin24}$ for A1 is not included as we can only determine a lower limit. This is because the conversion to T$_{\rm kin}$ exceeds the bounds of the rotation-to-kinetic temperature fit: the conversion factor from rotation to kinetic temperatures were fit  using collisional coefficients from $T_{\rm kin}$=0  to 300 K, with the fits then extrapolated out to $T_{\rm kin}$=500 K (\citealp{Ott2005} and \citetalias{Gorski2017}).

Similar to NGC\,253 there are two representative temperature components and there appears to be no spatial gradient in temperatures across the central kpc. In Figure \ref{fig:342TvDplot} we plot the kinetic temperatures as a function of offset from the dynamical center. The cool 27 K component is evenly distributed across the disk. The 308 K component also appears to be uniformly distributed across the disk, however the large uncertainties may hide a gradient, and in many locations we do not detect the \amm(4,4) line so the warm component is poorly sampled. \added{ In places where we can estimate an upper limit to $T_{\rm kin}$, the temperatures are consistent within the uncertainty to the flat temperature distribution. }

\added{It is  possible that the \amm\ molecule cannot survive along PDR surfaces as \amm\ is photodissociated  at $\sim$4.1 eV \citep{Suto1983} and hence does not sample the PDR component. This could account for the weaker detections we make towards C4 and C5 which are $\sim$30 pc from the dynamical center, while the other brighter emission regions C1-3 are $>$10\,pc further. From \citet{Meier2005} it appears that the A, B, and C clouds are largely affected by photo dissociation from the nuclear star cluster. A is dominated by photon PDRs while B and C are largely affected around the edges. To investigate this further we  calculate the molecular abundance of \amm\ (N$_{\rm NH_3}$/N$_{\rm H_2}$).  We accomplished this by smoothing our ammonia total flux (moment zero) maps to the resolution of the \aco\ map from \citet{Levine1994} ($\sim$2.7\arcsec).  We include only locations where the \amm(1,1) line could be reliably detected above 3$\sigma$ ($\sim$53\,K\,\kms) in the 2.7\arcsec\ resolution maps. Then, using a Galactic conversion factor from \citet{Strong1988} of X$_{\rm co}$=2.3$\times10^{20}$ cm$^{-2}$ (K \kms)$^{-1}$, we estimate the column density of \mh. The total \amm\ column density is estimated by extrapolating to include the \amm(0,0) state \citep{Ungerechts1986}. We use the rotational temperature derived earlier of 26 K, and the total \amm(1,1) column density (not just the upper inversion state) as described in \citet{Lebron2011}. The results are shown in Table \ref{tab:abundance}. We do not see evidence for abundance variations of \amm\ as the uncertainty is $\geq$35\% in all cases, given the rms of 12.5\,\added{K}\,\kms\ for the CO map and 17.8\,K\,\kms\ for the \amm(1,1) map. It follows that the PDRs do not preferentially destroy \amm\ molecules, and the lack of a temperature gradient across the C cloud suggests that PDR surfaces do not heat the gas on GMC scales.}
\explain{Edited paragraph on \amm\ abundances.}

\citet{Meier2011} \added{find that most of the dense gas in IC\,342 is concentrated in a dense cold component}. They analyze the dense gas in IC\,342 using HC$_3$N \added{and C$^{18}$O transitions}. They first attempt to explain the dense gas conditions with a uniform density and temperature. These models drastically over-predict the amount of dense gas in IC\,342  by 300\%$-$400\% for uniform cold dense gas with T$_{\rm kin}<$ 30 K.  In order to reproduce the observed masses derived from an optically thin isotopologue of CO, C$^{18}$O, they invoke a two-density component model of the gas.\added{ One low density component that emits only C$^{18}$O, and a high density low beam filling factor component that  emits C$^{18}$O and HC$_3$N}. For simplicity they assume that T$_{\rm kin}$ is the same for both components (30 K). They find that \replaced{most of the mass is}{the ratio of dense gas to total gas is (M$_{\rm dens} /$M$_{{\rm H}_2}$) $\sim$0.7} concentrated in a dense, low filling factor component. 
 We have measured two temperature components in the center of IC\,342, 27$\pm$3\,K and 308$\pm$171\,K, with an average, weighted by the average  column densities per location of the estimated \amm(0,0) and \amm(4,4) lines (respectively 3.0$\times10^{15}$\,cm$^{-2}$ and 6$\times10^{14}$ \,cm$^{-2}$), of 74$\pm$33\,K \added{assuming that the (4,4) line only comes from the warm component}. \replaced{ This suggests that a dense cold component doesn't completely dominate the mass distribution of IC\,342.}{If the cold dense component has a temperature of 27 K, then there must be more gas in the warm diffuse component to arrive at an average temperature of 74 K. For a single temperature of 70 K \citet{Meier2011} predict a dense gas fraction of $\sim$0.09}

Compared to NGC\,253 \citepalias{Gorski2017} there appears to be a larger difference between the cold and warm temperature components in IC\,342. The cool gas is cooler  (IC\,342: 27$\pm$3 K  vs. NGC\,253: 57$\pm$4 K) and the warm gas is the same temperature within the errors (IC\,342: 308$\pm$171\,K  vs. NGC\,253: 134$\pm$8\,K) . Both galaxies show an even distribution of temperatures across the central kpc. 

The kinetic temperatures in NGC\,2146 are only upper limits. Using a detection threshold of 20 K \kms\ for \amm(2,2) line, for the \replaced{A{\scriptsize I}}{A1} cloud we derive an upper limit of 89 K and for the \replaced{A{\scriptsize II}}{A2} cloud, 23 K. Without detections of the other \amm\ lines we cannot constrain the temperature of the gas any further.  There may also be a warmer component we cannot see.  More sensitive observations are necessary in order to derive temperatures via \amm\ thermometry. 

\added{Over the four galaxies in SWAN, two galaxies have sufficient detections of \amm\ metastable lines to perform thermometry. These two galaxies, NGC\,253 and IC\,342, show two temperature components, one cool and one warm. The other two galaxies, NGC\,6946 and NGC\,2146, have only one \amm\ metastable line detection and thus thermometry cannot be performed.}

\subsubsection{ Possible \amm(3,3) masers in IC\,342}
\added{There is a slight bump in the A1 and  C2 regions of IC\,342 (Figure \ref{fig:342select}) at the \amm(3,3) line. The positive slope of the Boltzmann plot suggests weak \amm(3,3) masers. This has been seen in NGC\,253 and NGC3079 ( \citealp{Ott2005}, \citealp{Miyamoto2015}, \citetalias{Gorski2017}), though it could also be due to a change in the ortho to para ratio of \amm.  The lowest energy state of ammonia is an ortho state, therefore the ortho to para ratio is sensitive to the formation temperature of the gas, with a colder formation temperature leading to more ortho species \citep{Takano2002}. In the Milky Way \amm(3,3) maser candidates are weakly associated with star forming sites (e.g., \citealp{Wilson1990} and \citealp{Goddi2015}). \citet{Meier2005} propose that the young stars forming in the C cloud are still enveloped in dense ``natal" material. These conditions could be appropriate for \amm(3,3) maser excitation in the C cloud of IC\,342.}  \explain{Moved section on \amm(3,3) masers here to better flow}

\subsubsection{ IC\,342 LVG Fitting with RADEX}

\begin{figure*}
\centering
\includegraphics[width=0.99\textwidth]{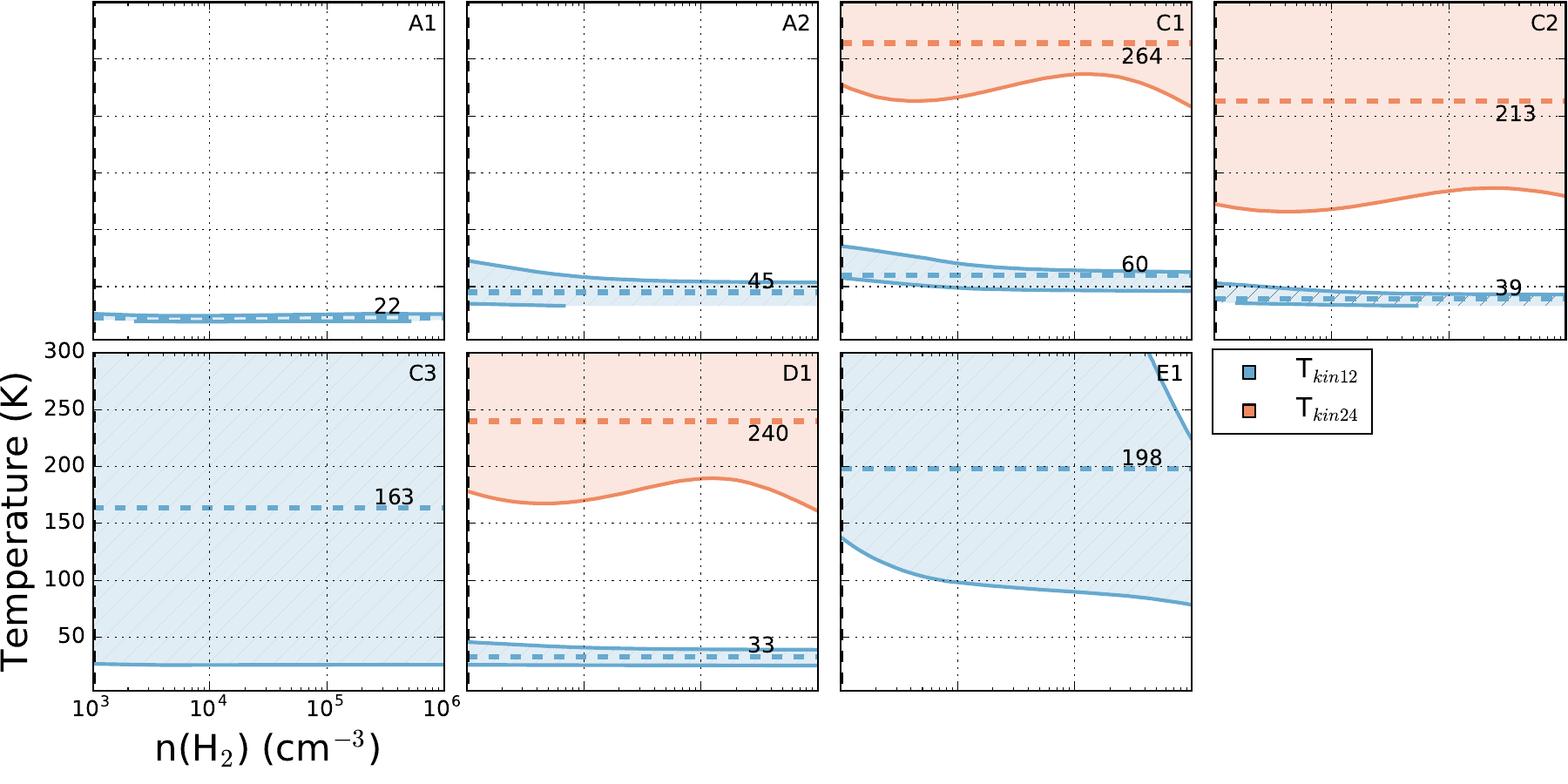}
\caption{RADEX LVG models based on the \amm\ lines in IC\,342. The blue region represents the fit to the \amm(1,1) to (2,2) ratio and the red region represents the fit to the \amm(2,2) to (4,4) ratio. The shaded area shows the $1\sigma$ confidence regions. The horizontal dashed line shows the median temperature fit for each instance.}
\label{fig:342lvg}
\end{figure*}

We can also perform direct fitting of LVG models in IC\,342 following the procedure from \citetalias{Gorski2017}. We perform this as a check of the approximations to LVG models described earlier. We use RADEX \citep{RADEX} with collisional coefficients from the LAMBDA database \citep{Schoier2005}. We generate a grid from 0 to 300 K in steps of 3 K and sample the collider density (H$_2$) logarithmically from 10$^3$ to 10$^6$ cm$^{-3}$ with 100 steps. The LVG approximation to radiative transfer accepts $N/\Delta v$ as the third axis (column density divided by line width). We use a line width of 34 \kms\ for IC\,342 and sample the column density from 10$^{13}$ to 10$^{17}$ cm$^{-2}$ with 100 steps.  The fits were carried out where we made detections of the \amm(1,1), (2,2) and (4,4) lines. The results are shown in Figure \ref{fig:342lvg}. Following \citetalias{Gorski2017} we plot the median temperature with a solid dashed line surrounded by 1$\sigma$ confidence contours. The cool component derived from the (1,1) to (2,2) ratio is shown in blue, and the warm component from the (2,2) to (4,4) line ratio is shown in red. The fits are unconstrained along the density axis (x-axis) in tune with \amm\ not being a good density probe. The fit to the (2,2) to (4,4) ratio is unconstrained along the temperature axis, therefore we can only provide a lower limit for those locations of $>$115 K. The sites C3 and E1 are also unconstrained along the temperature axis, therefore they are removed from the calculation of the average temperature from the (1,1) to (2,2) ratio. The average temperature, weighted by the larger of the asymmetric errors, is 27$\pm$2 K. These results are broadly consistent with the previous subsection.

\subsection{\water\ Masers}

Water masers span a large range of luminosities and can be variable on time scales of weeks (e.g., \citealp{Palagi1993}, \citealp{Claussen1996} and \citealp{Braatz1996}).    The water masers in all three galaxies are narrow, single peak spectral features with FWHM of $\sim$10 \kms.  In IC\,342 and NGC\,6946 the isotropic luminosities of the masers are of order 0.01 \Lsun. These are likely individual YSO or AGB stars (e.g., \citealp{Palagi1993}). The more luminous of the two masers seen in IC\,342 (W1) is located in the A cloud close to A1 and the other (W2) is located in the B cloud close to B1. The B cloud is a site of a young star forming region \citep{Meier2005}, supporting a possible origin of a YSO. The A cloud is a weaker site of star formation and is dominated by PDRs, so it is likely in a slightly more evolved state \citep{Meier2005}. Therefore the W2 maser may be either a stray YSO or an AGB star. Neither of these masers has been previously observed, though this is likely due to low sensitivity or variability ($\sim$10 mJy rms in \citealp{Tarchi2002}).

 In NGC\,6946 the maser is located in the southern spiral arm about 350 pc from the galaxy center. The luminosity is 0.042\,\Lsun, which is consistent with stellar maser rather than kilomaser luminosities.  Stellar \water\ masers are highly variable (e.g \citealp{Claussen1996}) and can be found in nuclear regions and spiral arms of galaxies. The lack of detections in the center of NGC\,6946 is thus unsurprising.   

NGC\,2146 has the most luminous masers in our sample of galaxies. All three of these masers have luminosities greater than 0.1 \Lsun\ classifying them as kilomasers.  Observed in June of 2001, masers W2 and W3 had luminosities of 0.5 \Lsun\ and 1.5 \Lsun\ respectively, while W1 was not detected \citep{Tarchi2002}. We observed these masers in July of 2013. We report isotropic luminosities  of 0.455 \Lsun\ and 2.087 \Lsun\ for 2146-W2 and 2146-W3 respectively, an increase of $\sim$33\% for 2146-W3, whereas 2146-W2 has the same flux as in 2002, but was variable before that \citep{Tarchi2002b}.  We calculate a luminosity of 0.351\,\Lsun\ for 2146-W1. The appearance of this maser indicates that it is variable. Including  the masers from NGC\,253 \citepalias{Gorski2017}, the galaxies with the most vigorous star formation host the kilomaser class of \water\ masers (253-W1, 2146-W2) while the lower luminosity stellar masers are more ubiquitous in the SWAN sample (253-W3, IC342-W1 and W2, and 6946-W1).

\subsection{\methanol\ Masers}

The 36\,GHz \methanol($4_{14}-3_{03}$)  maser was first detected in an extragalactic context by \citet{Ellingsen2014} in NGC 253.  These masers were resolved in \citetalias{Gorski2017} and evidence was given for correlation with weak shocks as the emission is morphologically similar to HNCO, which is unique compared to other molecular tracers \citep{Meier2015}. We have expanded the set of galaxies with 36\,GHz \methanol\ masers to include IC\,342 and NGC\,6946.

\begin{figure}
\includegraphics[width=0.49\textwidth]{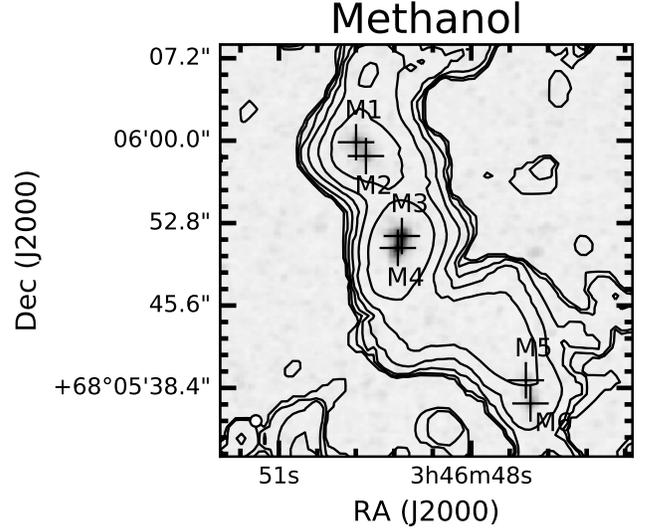}
\caption{ Peak flux map of the 36\,GHz methanol \added{maser} transition in greyscale. We show 3, 5, 8, 13, 21, and 34 times 50 mJy \kms\  contours are of the 96\,GHz \methanol(2$_k$-1$_k$) transitions from \citet{Meier2005}}
\label{fig:Meier_Methanol}
\end{figure}

There are six sites with  36\,GHz \methanol\ \added{maser} emission in IC\,342 coincident with the D$'$, C and E clouds. The \methanol\ emission is two orders of magnitude less luminous than the strongest emission found in NGC\,253 \citepalias{Gorski2017}. As shown in Figure \ref{fig:342select}, similar to NGC\,253, the morphology of the masers roughly matches that of HNCO emission mapped by \citet{Meier2005}, where they argue for weak shocks along the leading edges of the spiral arms. The morphological similarity to HNCO in NGC\,253 and IC\,342 implies that these two molecules are related.

 We compare with the thermal methanol lines at 96\,GHz from \citet{Meier2005} in Figure \ref{fig:Meier_Methanol}.  The similarity of the morphology of the 36\,GHz  lines with thermal lines is curious and might suggest that we are observing thermal 36\,GHz emission.  The strongest maser found in a survey of the inner 100 pc of the Milky Way by \citet{Yusef2013} (number 164) is 470 Jy\,\kms. The most luminous site in IC\,342 is M4 with a luminosity of  $\sim$0.08\,\Lsun. At the distance of the galactic center ($\sim$ 8\,kpc) this would equate to a 33,000 Jy\,\kms\ source, suggesting that this is indeed strong maser emission. Hence, if 36\,GHz \methanol\ is masing, as the brightness suggests, then both thermal and masing \methanol\ must trace elevated overall \methanol\ abundances, likely associated with weak shocks (e.g., \citealp{Meier2005})

There are two sites with 36\,GHz \methanol\ \replaced{emission}{masers} in NGC\,6946. These lie in the southern clump described in \citet{Schinnerer2006}. The two masers are an order of magnitude more luminous than those in IC\,342, and of similar luminosity to the masers in NGC\,253. It is possible that some portion of the emission is thermal, however because they are more luminous it is likely that these are masers.  The association with weak shocks in NGC\,253 and IC\,342 suggests that the southern clump is shocked by the inflow of material along the bar. 


\section{The Nature of Galaxy Nuclei from SWAN}

We have observed the central kpc of four star forming galaxies, NGC\,253, IC\,342, NGC\,6946, and NGC\,2146 in \amm, \water, and \methanol. These lines have shed light on the heating and cooling balance in the dense molecular ISM, the nature of class I masers (\water\ and \methanol) in an extragalactic context, and star formation feedback effects on 10's of pc scales. 


\subsection{ Heating and Cooling Balance}
Individual feedback effects  (e.g shocks, PDRs) do not appear to dominate the heating and cooling balance of the ISM on GMC scales. The uniformity of the two temperature distributions in NGC\,253 and IC\,342 show that shocks and PDRs do not cause variations in the temperature of the dense molecular ISM across kpc scales. \added{In \citetalias{Gorski2017} it is shown that there is no temperature correlation with enhancement of PDR tracer CN ($1-0$;$1/2-1/2$) in NGC\,253. In IC\,342 \citet{Meier2005} determine that clouds A, B, and C are affected by PDRs, through analysis of the molecular lines C$^{34}$S($2-1$) and C$_2$H($1-0,3/2-1/2$),  where no temperature correlation is observed.} It is important to note that the \amm\ molecule could be destroyed where feedback dominates (\amm\ is photodissociated  at $\sim$4.1 eV \citealt{Suto1983}), thus we would not be able to use \amm\ as a temperature probe.  The edge of the C cloud and the B cloud in IC\,342 may be good examples of this. We do not detect the edge of the C cloud closest to the PDRs  or the B cloud in \amm(1,1), (2,2) or (4,4). \amm(3,3) is a stronger emission line and more easily detected. We can see that there is \amm\ in the clouds but it may be depleted by ionizing photons, \added{though the \amm\ molecule is not preferentially dissociated as the abundance does vary significantly from the average value of ~12$\times$10$^{-9}$ N$_{\rm NH_3}$/N$_{\rm H_2}$  across the central kpc of IC\,342.} 

Should the \amm\ molecule survive in these environments it presents a picture where the heating and cooling balance of the molecular ISM is dominated by larger scale effects. Perhaps the distribution of temperatures is governed by cosmic rays or turbulent heating. In NGC\,253 the FWHM of the \aco\ line is more or less constant \added{($\sim$25\kms)} over the central kpc (e.g.\,\citetalias{Gorski2017}  and \citealt{leroy2015})   giving credence to the idea that the gas might be turbulently heated as the temperature is also evenly distributed. In IC\,342 the FWHMs of the \amm\ lines change drastically from cloud to cloud, from $\sim$20\,\kms\ in the A cloud to $\sim$60\,\kms\ in the D cloud. If turbulent heating is  a dominant factor we might naively assume that the D cloud would be warmer, however the temperatures remain roughly the same from cloud to cloud in IC\,342.  

\citet{Ginsburg2016} analyzed the temperatures of the dense molecular gas in the central 300 pc of the Milky Way using the H$_2$CO (formaldehyde) $3_{2,1}-2_{2,0}/3_{0,3}-2_{0,2}$ line ratio.  Here we use this study as a lens to understand our dense gas temperatures.  As they have only a single line ratio they do not measure two temperature components per location.  H$_2$CO also has a higher critical density than \amm, $\sim10^5$ cm$^{-3}$ compared to $\sim10^3$ cm$^{-3}$, meaning we are more sensitive to \replaced{more diffuse gas}{intermediate  densities.} They show that cosmic ray ionization rates (CRIRs) must be relatively low for cold gas ($\lesssim$ 60 K) to survive and that temperatures in the Milky Way can be explained by turbulent heating alone. They describe the temperature distribution as mostly uniform with temperatures 50-120 K, with temperatures generally increasing towards denser clouds. 

Nine of 13 galaxies show multiple temperature and velocity components in \citet{Mangum2013}. They used single pointings from the Green Bank Telescope to detect \amm\ lines.  The two component gas model appears to prevail in star forming galaxies.  These galaxies appear to have uniform distributions of temperatures (e.g \citealt{Ginsburg2016}, \citetalias{Gorski2017}, and this paper), though not all of these galaxies have been observed with GMC scale resolution. \citet{Meier2011} presents  evidence for two density components in IC\,342. Considering these studies we imagine two pictures. One, where our galaxies are dominated by turbulent heating, having diffuse (n $<10^5$ cm$^{-3}$) molecular gas temperatures less than 60 K and warmer dense clumps, and the CRIR  does not dominate the heating and cooling balance. Two, there is a cold dense component that cosmic rays cannot penetrate (e.g. \cite{Clark2013}) and a warm diffuse component (n $<10^5$ cm$^{-3}$) heated by cosmic rays.  \added{ The cosmic ray ionization rate in NGC\,253 is likely two orders of magnitude larger that that of the Milky Way ($\zeta_{\rm NGC\,253}=750\zeta_{Gal}\sim3\times10^{-14}\,{\rm s}^{-1}$) \citep{Bradford2003}, whereas in IC\,342 is constrained to be at least the  Milky Way value  $\zeta_{\rm IC\,342} > 10^{-17}$ s$^{-1}$ \citep{Meier2005}.} This may vary from cloud to cloud and galaxy to galaxy and would be revealed by resolving clump substructure \added{and by providing better constraints on cosmic ray ion-ionization rates.}

The difference in the two temperature components changes fairly drastically between galaxies. In NGC\,253, the difference between the cool and warm component is $\sim$77$\pm$12\,K and in IC\,342 it is $\sim$281\,$\pm$174K. Perhaps increased turbulence in NGC\,253 driven by the enhanced star formation \added{(NGC\,253's $\sim2$ \sfr\ star formation rate concentrated in 1 kpc compared to IC\,342's 2.8 \sfr\ global star formation rate)} mixes the two temperature components. Though if this were the case, then we might expect to see warmer gas in NGC\,2146 which has the highest star formation rate of our survey. This may also be a resolution effect: the larger the physical scale of the beam the more sensitive we are to diffuse gas (Figure 11 in \citealt{Ginsburg2016}).  More sensitive observations are needed to do proper \amm\ thermometry in NGC\,2146. 

\subsection{ Maser Luminosities and Star Formation}

\begin{figure}
\includegraphics[width=0.48\textwidth]{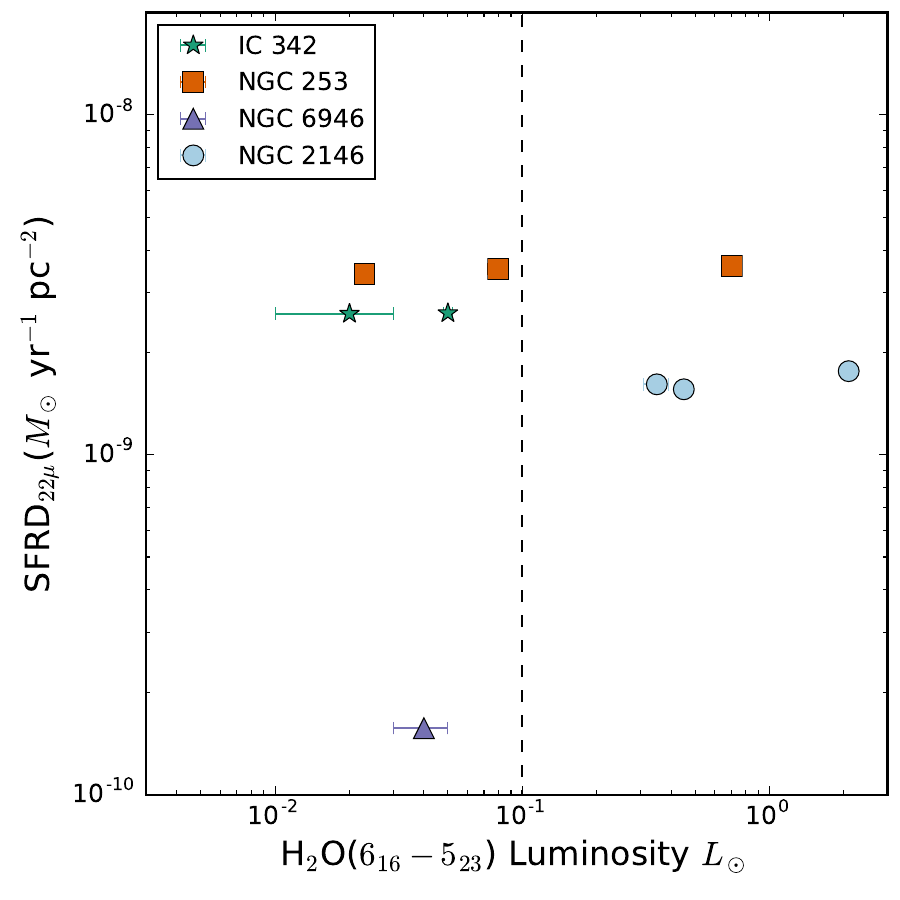}
\caption{\water\ maser luminosity versus 22$\mu$m SFR. The SFR is estimated from the corresponding pixel in the WISE Atlas images. The WISE data have a resolution of $\sim$12\arcsec. The uncertainty on the SFRD values is due to the flux calibration and  is 5\% \citep{Jarrett2013} and is smaller than the marker size. The vertical dashed line denotes the division between stellar and kilomaser classes. }
\label{fig:342water_all}
\end{figure}

In Figure \ref{fig:342water_all} we plot luminosities of all the water masers against the star formation rate surface density (SFRD), calculated from the corresponding pixel in the 22$\mu$m WISE Atlas images. \added{The star formation rate calibration uncertainty is $\pm 0.4$ \sfr} \citep{Jarrett2013}. A single WISE pixel represents the beam averaged emission at that location. The FWHM of the point spread function of the WISE Atlas images is 11.99\arcsec$\times$11.65\arcsec. We do not observe a general increase in \water\ maser luminosity with increased SFRD.  

  We observe two classes of masers, the stellar masers and the kilomasers. The kilomasers exist in the galaxies classified as starbursts. The kilomasers have luminosities $>$0.1\,\Lsun\ and can have many spectral components. They also correspond with the sources with larger star formation rates. The stellar masers are more evenly distributed across the sample of galaxies and have narrow single velocity components with luminosities $<$0.1\,\Lsun. Our results are consistent with the current classification of \water\ masers \citep{Hagiwara2001}. We do not observe any megamasers as none of our galaxies hosts an AGN.  In G17 we show evidence for a possible extension of water masers in NGC\,253 along the minor axis. The extension is likely due to star formation and not an AGN.

\begin{figure}
\includegraphics[width=0.49\textwidth]{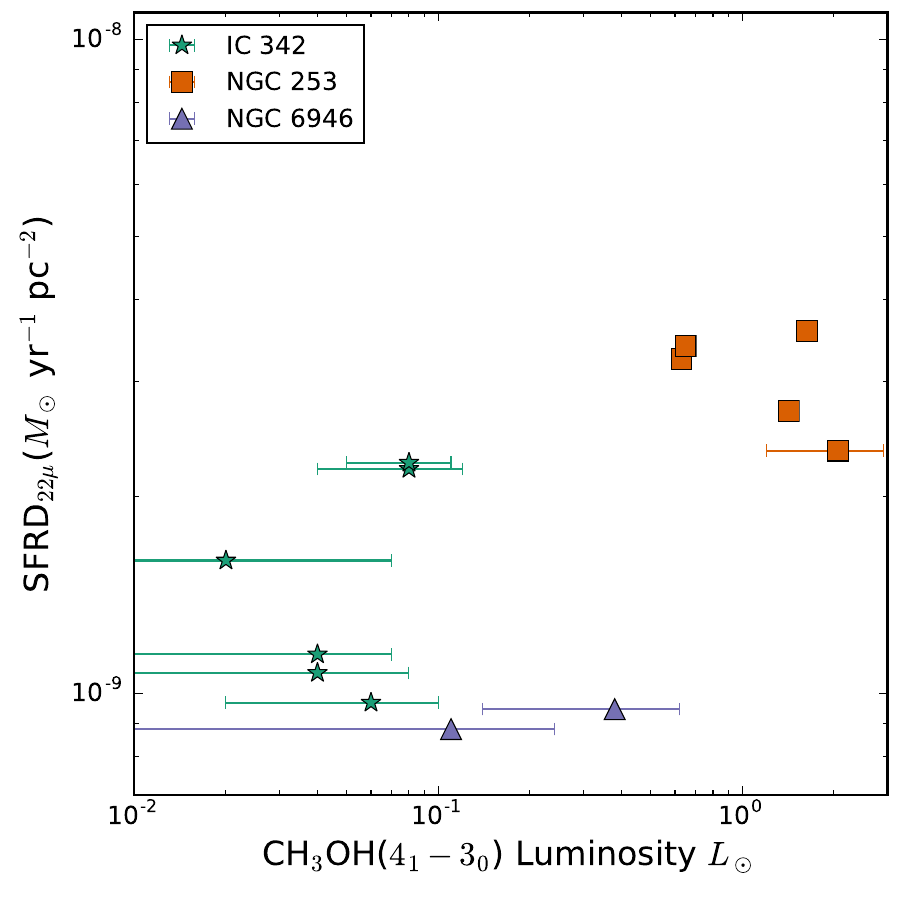}
\includegraphics[width=0.49\textwidth]{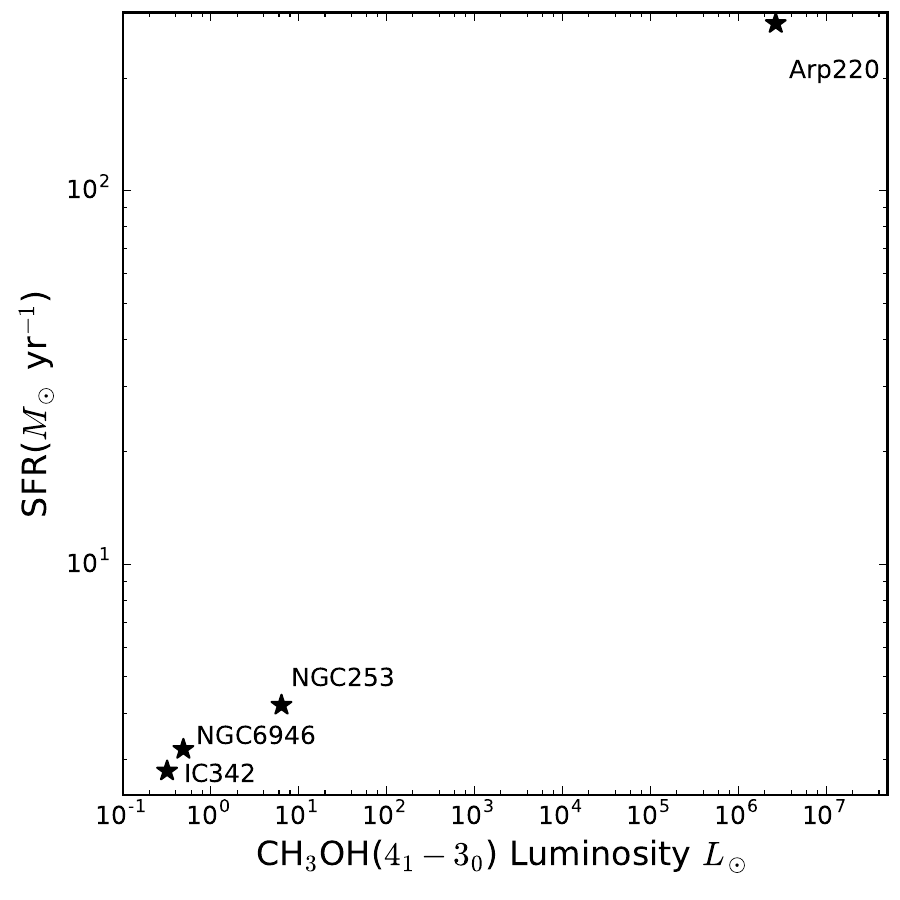}
\caption{Top: total \methanol\ maser luminosity versus 22$\mu$m SFRD estimated from the corresponding pixel in the WISE Atlas images. The WISE data have a resolution of $\sim$12\arcsec. The uncertainty on the SFRD values is 5\% \citep{Jarrett2013}. Bottom: galactic star formation rate from \citet{Gao2004} versus total \methanol\ maser luminosity for all extragalactic sources with 36\,GHz \methanol\ detections to date.}\label{fig:methanol_all}
\end{figure}

In the three extragalactic cases where 36\,GHz \methanol\ \added{maser} emission was targeted it was detected. The ubiquitous detections open up a new possibility for probing shocked environments in the extragalactic context. As shown in Figure \ref{fig:methanol_all}, the luminosity of an individual maser does not apparently scale with local SFRD. However, as shown in Figure \ref{fig:methanol_all},  for all galaxies external to the Milky Way with detected 36\,GHz \replaced{emission}{masers}, we plot the global star formation rate versus the total 36\,GHz \methanol\ luminosity. There appears to be a close relationship between  the total luminosity of the 36\,GHz \added{maser} line and the global star formation rate, albeit we have only four galaxies. This is perhaps unsurprising as these galaxies have large reservoirs of molecular gas, but the result also suggests little variability of the masers. 

 \citet{Yusef2013}, \citet{Ellingsen2014} and \citet{Chen2015} discuss the possibility that cosmic rays release \methanol\ into the ISM resulting in higher abundances towards a galaxy's center as the cosmic ray density increases, however if the cosmic ray density is too high \methanol\ molecules will be destroyed. Should the cosmic ray density be highest in the center of galaxies, the concentration of masers at the edge of the molecular bar in NGC\,253 (\citealp{Ellingsen2014} and \citetalias{Gorski2017}) and IC\,342 would be consistent with this idea. However in Arp\,220  \citet{Chen2015} observe a \methanol\ X-ray correlation. They take X-ray emission to reflect the cosmic ray density. If ture, this would suggest that \methanol\ is not destroyed in regions where the cosmic ray density is high. At this point then, this idea is difficult to test via extragalactic \methanol\ observations. 
 
 The similar morphology of the HNCO emission and \methanol\ masers in NGC\,253 and IC\,342  further supports a different generating mechanism as the HNCO molecule appears to be anti-correlated with PDRs. \citet{Meier2005} and \citet{Meier2015} suggest that weak shocks and/or ice mantle evaporation are related to the HNCO and \methanol\ emission. In IC\,342 they suggest the emission is reflective of shocks along the leading edge of the molecular bar. Because of the morphological similarities between HNCO and 36\,GHz \methanol \added{masers}, the physical conditions that give rise to emission are likely related. In Arp220 the \methanol\ is correlated with the X-ray plume generated by a starburst super-wind \citep{Chen2015}. It is possible that the \methanol\ is related to the shocks driven by the wind. In \citetalias{Gorski2017} we suggest the \methanol\ masers are related to shocks in expanding superbubbles.  So far, a relationship between 36\,GHz \methanol\ \added{maser} emission with weak shocks would be a more consistent picture across all galaxies.
 
 It is not yet known what fraction of the emission in these galaxies is thermal vs. non-thermal (maser). It's possible there are multiple conditions that give rise to 36\,GHz \added{\methanol} emission or that separate conditions give rise to thermal versus maser emission.   More in-depth studies are critical to uncover the pumping of the maser and the nature of the 36\,GHz \methanol\ line in the extragalactic context.  
 

\section{Summary}
With this paper we complete the analysis of the \water, \amm, and \methanol\ lines in the Survey of Water and Ammonia in Nearby galaxies (SWAN).   The primary results are:

\begin{enumerate}

\item We have detected metastable \amm\ transitions in IC\,342, NGC\,6946, and NGC\,2146.  In IC\,342 the two molecular spiral arms and the central molecular ring are traced in \amm(1,1) to (4,4). We make only one detection of the \amm(3,3) line in NGC\,6946 near the center \added{ and two weak detections of the \amm(1,1) line in NGC\,2146.} 

\item \added{Of the four SWAN galaxies we were able to perform \amm\ thermometry over NGC\,253 and IC\,342. Individual feedback effects (e.g. supernovae, PDRs, shocks) do not appear to dominate the distribution of temperatures. The two temperature components in both galaxies have flat distributions across their central molecular zones. Compared to NGC\,253,  the molecular gas components in IC\,342 have a larger temperature difference, as the cold component is colder. }

\item We detect a uniform 27$\pm$3 K dense molecular gas component across the central kpc of IC\,342. We also detect a 308$\pm$171 K component with indications of a uniform distribution. The direct LVG models are consistent with our LVG approximation. The temperatures of the clouds do not appear to be affected by the nuclear PDRs given the flat temperature distribution across the C cloud. We also provide evidence for weak \amm(3,3) masers in the A and C clouds. The dense molecular gas in NGC\,2146 must be fairly cold, with T $<89$ K, and there is no evidence for a hot component. 

\deleted{Compared to NGC\,253,  the molecular gas components in IC\,342 have a larger temperature difference, as the cold component is colder. The two temperature components in both galaxies have flat distributions across their central molecular zones. Individual feedback effects (e.g. supernovae, PDRs, shocks) do not appear to dominate the distribution of temperatures.}\explain{moved to item 2}

\item We have detected two new stellar \water\ masers in IC\,342, one in NGC\,6946, and three \water\ kilomasers in NGC\,2146 of which one was previously undetected. Across the entire SWAN sample the kilomasers are found in the two starburst galaxies NGC\,253 and NGC\,2146, while the lower luminosity stellar masers are found more uniformly across the sample. 

\item  We report the first detection of \added{the} 36\,GHz \methanol\ \added{line} in IC\,342 and NGC\,6946. This expands the number of galaxies beyond the Milky Way with 36\,GHz \methanol\ emission to four including Arp220 \citep{Chen2015}. The morphology in IC\,342 and NGC\,253 is similar to HNCO emission, implying that weak shocks may pump the maser. It is possible there is a mixture of thermal and non-thermal (maser) emission that high resolution observations will separate. The luminosity of the emission appears to roughly scale with the global strength of the star formation activity. 

\end{enumerate}

\acknowledgments 
Mark Gorski acknowledges support from the National Radio Astronomy Observatory in the form of a graduate student internship, and a Reber Fellowship. This research has made use of the NASA/IPAC Extragalactic Database (NED) and NASA/ IPAC Infrared Science Archive, which is maintained by the Jet Propulsion Laboratory, Caltech, under contract with the National Aeronautics and Space Administration (NASA) and NASA's Astrophysical Data System Abstract Service (ADS).

\software{CASA \citep{mcmullin07}, RADEX \citep{RADEX}}

\clearpage


\begin{deluxetable*}{llllllllll}
\tabletypesize{\footnotesize}
\tablewidth{0pt}
\tablecolumns{8}
\tablecaption{\amm\ Line Parameters \label{tab:342amm}}
\tablehead{ 
\colhead{Source} 	&\colhead{Transition} 	& \colhead{RA (J2000)}& \colhead{DEC (J2000) }			& \colhead{$\int T_{\rm mb} d\nu$} 	& \colhead{$V_{\rm LSRK}$ }	& \colhead{$V_{\rm FWHM}$} 	& \colhead{$T_{\rm mb}$} &  \\
 				&			& {hh}:{mm}:{ss} 	&{\arcdeg}~\phn{\arcmin}~\phn{\arcsec}	&(K \kms)					&(\kms)				&(\kms)					&(K)	
}
\startdata
\cutinhead{IC\,342 }
A1 			& (1,1)& 03:46:48.6	& 68:05:43.4 & \phn70.5$\pm$5.0 	& 21.6$\pm$2.1 	& 32.5$\pm$4.9 	& 2.0$\pm$0.3\\
			& (2,2)& 	&	& \phn22.3$\pm$3.8	& 26.8$\pm$1.6	& 12.5$\pm$3.9	& 1.7$\pm$0.5\\
			& (3,3)& 	&	& \phn86.7$\pm$4.3	& 21.8$\pm$1.5	& 28.1$\pm$3.5	& 2.9$\pm$0.2\\
			& (4,4)& 	&	& \phn22.1$\pm$3.9	& 18.6$\pm$1.9	& 14.0$\pm$4.4	& 1.5$\pm$0.4\\  \hline
A2			& (1,1)& 03:46:48.4	& 68:05:43.0 & \phn71.1$\pm$6.7	& 26.2$\pm$3.3	& 43.5$\pm$7.8	& 1.5$\pm$0.2\\
			& (2,2)&	& 	& \phn37.2$\pm$7.1	& 22.3$\pm$2.8	& 27.4$\pm$8.9	& 1.3$\pm$0.4\\
			& (3,3)&	& 	& \phn67.6$\pm$5.5	& 19.0$\pm$2.4	& 32.0$\pm$5.6	& 2.0$\pm$0.3\\ \hline
B1			& (3,3)& 	03:46:47.8		& 68:05:46.4	& \phn27.5$\pm$3.7		& 25.4$\pm$1.6	& 13.0$\pm$3.4	& 2.0$\pm$0.5\\ \hline
C1 			& (1,1)&	03:46:49.0	& 68:05:51.7 	& \phn98.7$\pm$7.6	& 51.3$\pm$3.4	& 51.0$\pm$8.1	& 1.8$\pm$0.3\\
			& (2,2)&	& 	& \phn65.0$\pm$4.6	& 51.0$\pm$1.9	& 30.2$\pm$4.5	& 2.0$\pm$0.3\\
			& (3,3)& 	& 	& 139.0$\pm$4.5	& 47.5$\pm$1.3	& 33.6$\pm$3.1	& 3.9$\pm$0.3\\
			& (4,4)&	& 	& \phn52.8$\pm$7.1	& 44.8$\pm$4.0	& 42.0$\pm$9.4	& 1.2$\pm$0.2\\ \hline
C2		 	& (1,1)&	03:46:49.2 	& 68:05:49.9	&103.3$\pm$6.7	& 45.8$\pm$2.8	& 47.2$\pm$6.7	& 2.1$\pm$0.3\\
			& (2,2)&	& 	& \phn51.1$\pm$5.5	& 45.4$\pm$2.5	& 27.4$\pm$5.9	& 1.8$\pm$0.3\\
			& (3,3)&	& 	&138.0$\pm$3.6	& 50.4$\pm$1.1	& 35.3$\pm$2.6	& 3.7$\pm$0.2\\
			& (4,4)&	& 	& \phn34.4$\pm$9.9	& 53.5$\pm$7.8	& 50.9$\pm$18.3	& 0.6$\pm$0.2\\ \hline
C3			& (1,1)&  	03:46:49.1 	& 68:05:48.0 & \phn84.9$\pm$5.5	& 49.0$\pm$2.7	& 50.0$\pm$6.3	& 1.6$\pm$0.2\\
			& (2,2)&	& 	& \phn55.8$\pm$7.2	& 42.8$\pm$3.6	& 35.1$\pm$8.5	& 1.5$\pm$0.3\\
			& (3,3)&	& 	& \phn77.6$\pm$4.1	& 49.8$\pm$1.5	& 28.2$\pm$3.5	& 2.6$\pm$0.3\\ \hline
C4			& (3,3)&		03:46:49.0 	& 68:05:46.3 	& \phn25.2$\pm$3.4		& 53.8$\pm$1.5	& 14.8$\pm$3.4	& 1.6$\pm$0.3  \\ \hline
C5			& (3,3)&   	03:46:48.9	& 68:05:48.7 	&  \phn57.7$\pm$6.0	& 57.9$\pm$3.3	& 42.8$\pm$7.7	& 1.3$\pm$0.2 \\ \hline
D$'$1		 	& (1,1)&	03:46:49.8	& 68:05:59.7	& 108.4$\pm$16.4	& 61.6$\pm$8.7	& 90.8$\pm$22.9	& 1.1$\pm$0.2\\
			& (2,2)&	& 	& \phn42.0$\pm$8.8	& 50.8$\pm$5.2	& 36.7$\pm$12.2	& 1.1$\pm$0.3\\
			& (3,3)&	& 	& \phn69.4$\pm$5.5	& 47.7$\pm$2.1	& 27.1$\pm$4.9	& 2.4$\pm$0.4\\
			& (4,4)& 	& 	& \phn36.9$\pm$9.5	& 61.9$\pm$6.3	& 39.9$\pm$14.8	& 0.9$\pm$0.3\\ \hline
D$'$2		 	& (1,1)&	03:46:49.7	& 68:05:58.4 	& \phn89.4$\pm$8.1	& 49.0$\pm$4.1	& 57.7$\pm$9.8	& 1.5$\pm$0.2\\
			& (3,3)&	& 	& \phn52.8$\pm$5.4	& 54.0$\pm$2.3	& 25.2$\pm$5.4	& 2.0$\pm$0.4\\ \hline
D$'$3			& (1,1)&	03:46:49.4	& 68:06:02.0 	& \phn29.1$\pm$2.3	& 23.7$\pm$0.9	& 11.0$\pm$1.9		& 2.5$\pm$0.4\\ \hline
E1		 	& (1,1)&	03:46:47.5 	& 68:05:42.7	& \phn51.9$\pm$5.6	& 18.3$\pm$2.5	& 27.9$\pm$6.0	& 1.7$\pm$0.3\\
			& (2,2)&	& 	& \phn53.2$\pm$6.6	& 21.0$\pm$3.4	& 34.3$\pm$7.9	& 1.5$\pm$0.3\\
			& (3,3)&	& 	& \phn47.3$\pm$3.3	& 12.9$\pm$1.1	& 15.5$\pm$2.5	& 2.9$\pm$0.4\\ \hline
E3			& (3,3)&	 03:46:47.0	& 68:05:37.1	& \phn28.0$\pm$3.2	& 25.3$\pm$1.4	& 13.5$\pm$2.0	& 1.9$\pm$0.4\\ \hline
\cutinhead{NGC\,6946 }
A{\scriptsize I} 	& (3,3)& 20:34:52:9	&	60:09:14.4 	& \phn21.8$\pm$3.8	& 7.9$\pm$2.7		& 26.9$\pm$6.4	& 0.8$\pm$0.2\\ \hline
\cutinhead{NGC\,2146}
A{\scriptsize I} 	& (1,1)&	06:18:37.2	&78.21.25.1 	&\phn17.2	$\pm$0.8	&944.3$\pm$1.0	&6.5$\pm$1.7		& 2.5$\pm$0.6\\ 
A{\scriptsize II} 	& (1,1)&	06:18:35.1	&78.31.34.1 	&\phn44.7	$\pm$5.9	&916.3$\pm$3.7	&39.3$\pm$8.7		&1.1$\pm$0.2  \enddata
\end{deluxetable*}


\begin{deluxetable*}{llllllllll}
\tabletypesize{\footnotesize}
\tablewidth{0pt}
\tablecolumns{8}
\tablecaption{\water\  Line Parameters \label{tab:342water}}
\tablehead{ 
\colhead{Source} 	& \colhead{RA (J2000)}& \colhead{DEC (J2000) }			& \colhead{$\int T_{\rm mb} d\nu$} 	& \colhead{$V_{\rm LSRK}$ }	& \colhead{$V_{\rm FWHM}$} 	& \colhead{$T_{\rm mb}$} &  Luminosity\\
 				& {hh}:{mm}:{ss} 	&{\arcdeg}~\phn{\arcmin}~\phn{\arcsec}	&(K \kms)					&(\kms)				&(\kms)					&(K)				& L$_{\odot}$
}
\startdata
\cutinhead{IC\,342 }
IC342-W1 	& 03:46:48.7 	& 68 05 43.7	& 97.9$\pm$32.8	&	26.5$\pm$0.2	&	13.0$\pm$0.5		&	7.0$\pm$0.2	& 0.017$\pm$0.006	\\  
IC342-W2	$^*$	& 03:46:47.7	& 68 05 45.9 	& 29.1$\pm$11.0	&	20.0$\pm$1.0	&	\phn9.1$\pm$2.7	&	3.0$\pm$0.3	& 0.005$\pm$0.003	\\ 
\cutinhead{NGC\,6946 }
6946-W1	& 20:34:52.9	& 60 08 51.3 	& 58.3$\pm$2.2	&	94.8$\pm$0.6	&	11.0$\pm$1.1	&	5.0$\pm$0.4	&	0.042$\pm$0.002\\
\cutinhead{NGC\,2146}
2146-W1	& 06:18:38.7	& 78 21 19.7	& \phn72.6$\pm$9.3	&	\phn831.4$\pm$0.8	&	\phn4.5$\pm$1.3	&15.2$\pm$4.0		&	0.351$\pm$0.045 \\
2146-W2	& 06:18:38.7	& 78 21 24.3	& \phn94.1$\pm$3.8	&	\phn831.9$\pm$0.5	&	\phn9.7$\pm$1.1	&\phn9.1$\pm$0.9	&	0.455$\pm$0.018 \\
2146-W3	& 06:18:36.6	& 78 21 27.6	& 431.5$\pm$7.6	&	1013.3$\pm$1.4	& 	48.0$\pm$3.4		&\phn8.4$\pm$0.5	&	2.087$\pm$0.037 \\
\enddata
\tablecomments{$^*$ not deconvolved from the spectral resolution}
\end{deluxetable*}


\begin{deluxetable*}{llllllllll}
\tabletypesize{\footnotesize}
\tablewidth{0pt}
\tablecolumns{8}
\tablecaption{\methanol\ Line Parameters \label{tab:342methanol}}
\tablehead{ 
\colhead{Source} 	& \colhead{RA (J2000)}& \colhead{DEC (J2000) }	& \colhead{$\int T_{\rm mb} d\nu$} 	& \colhead{$V_{\rm LSRK}$ }	& \colhead{$V_{\rm FWHM}$} 	& \colhead{$T_{\rm mb}$} &  Luminosity\\
 				& {hh}:{mm}:{ss} 	&{\arcdeg}~\phn{\arcmin}~\phn{\arcsec}	&(K \kms)				&(\kms)				&(\kms)					&(K)		&\Lsun		
				}
\startdata
\cutinhead{IC\,342 }
IC\,342-M1 	& 03:46:49.8 	& 68 05 59.9	& 140.4$\pm$93.6		&	51.9$\pm$0.8	&	24.2$\pm$1.9	&	5.5$\pm$0.4	&	0.061$\pm$0.041	\\  
IC\,342-M2	& 03:46:49.6	& 68 05 58.7 	& \phn94.5$\pm$91.7	&	54.5$\pm$0.9	&	19.7$\pm$2.1	&	4.5$\pm$0.4	&	0.041$\pm$0.039	\\ 
IC\,342-M3	& 03:46:49.1	& 68 05 51.7 	& 176.9$\pm$83.3		&	49.9$\pm$0.6	&	23.5$\pm$1.5	&	7.1$\pm$0.4	&	0.077$\pm$0.036 	\\
IC\,342-M4	& 03:46:49.1	& 68 05 50.6 	& 186.5$\pm$60.5		&	50.6$\pm$0.5	&	23.2$\pm$1.1	&	7.6$\pm$0.3	&	0.081$\pm$0.026	\\
IC\,342-M5	& 03:46:47.1	& 68 05 39.1 	& \phn43.0$\pm$118.8	&	21.3$\pm$1.1	&	11.6$\pm$3.1	&	3.5$\pm$0.7	&	0.019$\pm$0.052	\\
IC\,342-M6	& 03:46:47.1	& 68 05 37.0 	& \phn83.2$\pm$77.9	&	23.1$\pm$0.8	&	16.6$\pm$1.7	&	4.7$\pm$0.4	&	0.036$\pm$0.033	\\
\cutinhead{NGC\,6946 }
6946-M1 	& 20:34:52.8 	& 60 09 13.4	& 269.6$\pm$149.7		&	33.3$\pm$1.5	&	56.1$\pm$3.5	&	4.5$\pm$0.2	&	0.376$\pm$0.208	\\  
6946-M2 	& 20:34:52.7 	& 60 09 11.2	& \phn82.3$\pm$98.3	&	66.0$\pm$1.1	&	22.4$\pm$2.6	&	3.5$\pm$0.4	&	0.115$\pm$0.137	\\  
\enddata
\end{deluxetable*}


\begin{deluxetable*}{llllllll}
\tabletypesize{\footnotesize}
\tablewidth{0pt}
\tablecolumns{5}
\tablecaption{\amm\ Temperatures \label{tab:342temp}}
\tablehead{ 
\colhead{Source} 		& \colhead{$T_{12}$} 	& \colhead{$T_{24}$}	& \colhead{$T_{\rm kin12}$}	& \colhead{$T_{\rm kin24}$}  \\
 					&(K)					&(K)					&(K)					&(K)	
}
\startdata
\cutinhead{IC\,342 }
A1		& 21$^{+3}_{-2}$	& 171$^{+133}_{-52}$	& 22$^{+4}_{-3}$	& $>$259\\
A2		& 29$^{+7}_{-5}$	& $<$ 96				& 34$^{+17}_{-9}$	& $<$170		\\
C1		& 34$^{+5}_{-4}$	&137$	^{+34}_{-24}$	& 48$^{+16}_{-10}$	& 363$^{+330}_{-132}$	\\
C2		& 27$^{+4}_{-3}$	&115	$^{+52}_{-31}$		& 32$^{+8}_{-5}$	& $<$240		\\
C3		& 34$^{+6}_{-5}$	& $<$ 75				& 48$^{+22}_{-12}$	& $<$113	\\
D$'$1 		&24$^{+6}_{-4}$	&148	$^{+150}_{-51}$	& 26$^{+11}_{-6}$	& 452$^{+7410}_{-281}$ \\
D$'$2		& $<$ 18			&					&$<$28	\\
D$'$3		& $<$ 35			&					&$<$51	\\
E1		& 53$^{+22}_{-12}$	& $<$ 77				& 154$^{+458}_{-81}$	& $<$126 \\
\cutinhead{NGC\,2146 }
A{\scriptsize I}		& $<$ 44			&					&$<$89	\\
A{\scriptsize II}		& $<$ 21			&					&$<$23	\\
\enddata
\end{deluxetable*}


\begin{deluxetable*}{llll}
\tabletypesize{\footnotesize}
\tablewidth{0pt}
\tablecolumns{5}
\tablecaption {\amm\ Abuncance\label{tab:abundance}}
\tablehead{
 Location	& N$_{H_2}$	& 	N$_{\rm NH_3}$					& Abundance 				\\
 		&	10$^{22}$(cm$^{-2}$)	&	10$^{14}$(cm$^{-2}$)		& 10$^{-9}$(N$_{\rm NH_3}$/N$_{\rm H_2}$) 
}
\startdata
A1	&	14$\pm$3	&	13$\pm$6	&	\phn9$\pm$4	\\
A2	&	15$\pm$3	&	18$\pm$6	&	12$\pm$4	\\
B1	&	11$\pm$3	&	13$\pm$6	&	11$\pm$5	\\
C1	&	17$\pm$3	&	15$\pm$6	&	\phn8$\pm$3	\\
C2	&	15$\pm$3	&	20$\pm$6	&	14$\pm$4	\\
C3	&	12$\pm$3	&	13$\pm$6	&	12$\pm$5	\\
C4	&	12$\pm$3	&	...	&	...	\\
C5	&	\phn9$\pm$3	&	\phn9$\pm$6	&	10$\pm$6	\\
D1	&	\phn7$\pm$3	&	14$\pm$6	&	21$\pm$9	\\
D2	&	\phn8$\pm$3	&	16$\pm$6	&	21$\pm$8	\\
D$'$3	&	\phn7$\pm$3	&	13$\pm$6 &	18$\pm$8	\\
E1	&	15$\pm$3	&	\phn5$\pm$6	&	\phn3$\pm$4\\
E2	&	\phn7$\pm$3	&	...	&	...			
\enddata
\tablecomments{Molecular abundance of \amm. Note that these measurements are made with a $\sim$2.7\arcsec\ beam and therefore \deleted{the}not all locations are spatially independent.}
\end{deluxetable*}

\end{document}